# Low-skilled Occupations Face the Highest Upskilling Pressure

**Authors:** Di Tong[1], Lingfei Wu[2], James A. Evans[3,4]*

**Affiliations:**

[1] Sloan School of Management, Massachusetts Institute of Technology, Cambridge, MA 02142.

[2] School of Computing and Information, University of Pittsburgh, Pittsburgh, PA 15260.

[3] Department of Sociology, University of Chicago, 1126 E 59th St, Chicago, IL 60637.

[4] Knowledge Lab, University of Chicago, 5735 South Ellis Avenue, Chicago, IL 60637.

*Corresponding author. E-mail: jevans@uchicago.edu (J.E.)

**Substantial scholarship has estimated the susceptibility of jobs to automation, but little has examined how job contents evolve in the information age as new technologies substitute for tasks, shifting required skills rather than eliminating entire jobs. Here we explore patterns of occupational skill change and characterize occupations and workers subject to the greatest re-skilling requirements. Recent work found that changing skill requirements are greatest for STEM occupations in the 2010s. Nevertheless, analyzing 167 million online job posts covering 727 occupations, we find that skill change is greatest for low-skilled occupations when accounting for distance between skills. We further investigate the differences in skill change across employer and market size, as well as social demographic groups. We find that jobs from small employers and markets experienced larger skill upgrades to catch up with the skill demands of their large employers and markets. Female and minority workers are disproportionately employed in low-skilled jobs and face the most significant skill adjustments. While these varied skill changes could create uneven reskilling pressures across workers, they may also lead to a narrowing of gaps in job quality and prospects. We conclude by showcasing our model's potential to chart job evolution directions using skill embedding spaces.**

# Main

With rapid advances in automation, computerization, and Artificial Intelligence (AI) technologies, substantial scholarship and public debate has focused on estimating the susceptibility of jobs to full or partial automation and the employment consequences that follow[1–8]. By contrast, relatively little work has examined how job contents evolve. New technologies often substitute for some skills within jobs while simultaneously creating new ones, which alter job skill demands and impose re-skilling pressure on workers[9–12]. Here we focus on the future of work not in terms of job elimination and displacement, but job transformation[11,13].

Measuring job transformation by skill change is a relatively recent and emerging research program. The bulk of prior empirical studies focus on analyzing specific jobs, skills and industrial forces of change[9,11,14–17]. This work demonstrates the importance of compositional skill change, but lacks quantification of the similarity between skills within jobs. This makes it difficult to systematically compare jobs and understand which group of workers experience more skill change. The major challenge for quantifying systemic job skill change lies in obtaining datasets that accurately represent job skill requirements[13]. Recent research has shifted from inferring skill change indirectly from shifts in relative wages and the supply of highly-educated labor[18], to direct measurement of skill composition. Still, this work relies on static, coarse skill taxonomies such as the Dictionary of Occupational Titles[14], the Occupational Information Network Database (O*NET)[13,19], or job titles from Census coding volumes[10]. Most recently, Deming and Noray (2020)[20] developed the first direct, dynamic and precise measurement of job skill change using high-quality job advertisement data. They quantify skill changes for a job over time by adding up the adjustments in the frequency or emphasis of all skills, including those that became outdated and new ones that were required. For the detailed equation, see the Supplementary Information section 2.1.

The limitation of this quantification, however, lies in its inadequate consideration of the graduated relationship between different skills. This oversight results in a skewed perspective, inducing an upward bias for technology-intensive and specialized occupations that possess many more listed “skills” than occupations with less specialization. For these high-skilled and specialized occupations, when old skills are replaced by new ones, they are usually quite similar and likely not too difficult to learn, similar to how learning a new programming language is more straightforward for someone who already knows several. That measurement approach to skill change also exhibits downward bias for low-skilled occupations, where new skills needed are fewer yet substantially different and more challenging to acquire. To overcome these limitations, we propose to investigate and quantify the “distance” between skills within the underlying space of complex human capital[21]. Some skills are closer to one another, creating local, accessible transition paths. This type of transition also implies less radical changes to the nature of the job, with smaller impacts on its returns and prospects. As illustrated by Fig.1a-b, the transition required for a computer programmer to adapt to a shift in demand within the “IT & Software Development” skill cluster from one coding language to another is smaller than the required shift for a food batchmaker who manufactures large quantities of food for packaging and distribution to their first coding language (database administration).

Given this concern, we provide an alternative quantification method of skill change that uses embedding distance to model the space between skills. Deming and Noray (2020) suggested that high-skilled STEM and technology-intensive occupations experienced the most skill content change from 2007 to 2019[20]. Here we use the same dataset as Deming and Noray[20], but present a dramatically different observation when accounting for distance between skills: low-skilled occupations changed *much* more than high-skilled occupations, suggesting very different employment and skilling policies. Using the skill probability change measure as suggested by Deming and Noray[20], computer programmers change 175% more than food batchmaker (7.453 vs 2.708) (Fig.1c). After modeling skill distance using skill vectors learned from direct and indirect co-occurrences of skills in job ads with word embedding[22], however, the assessment of who experiences the most radical skill change is reversed—food batchmaker changes 144% more than computer programmer (0.082 vs. 0.017 in cosine distance; Fig.1d). We note that an intermediate version of Deming and Noray's (2020) measure, formally equivalent to their original metric[20] but using data-driven clusters also reverses their assessment (see Supplementary Information section 2.3).

The Methods section explains how we quantify skill change by recovering skill distance with a machine learning model[22] that embeds skills in a high-resolution skill space. Based on this improved measurement, we analyze skill requirements from 727 occupations across 167 million job ads, which cover the near-universe of the U.S. labor market in the past decade. Our findings confirm that occupations categorized as low-skilled encounter the most radical skill upgrades. This holds true regardless of the criteria used for distinguishing between low and high-skill roles, whether by skill number, pay level, or educational degree (Fig.2). To better understand skill change variations beyond low and high skill jobs, we further compare occupational skill change across employer and market sizes. Our evidence suggests that small labor markets and small employers change more relative to their larger counterparts as they upgrade their skills and converge to the skill demand of large markets and employers[23] (Fig.3). This "catching-up" pattern indicates that skill changes are more incremental at the frontier, where most high-skilled occupations, large employers, and markets are concentrated. In contrast, low-skilled occupations, small employers, and markets require larger leaps in the skill space as they catch up. Additionally, we probe into the implication of varying skill change across occupations on workers of different demographic groups. Female and minority workers are disproportionately distributed in low-skilled jobs and experience more significant up-skilling requirements than male and white workers (Fig.3). We conclude by discussing the broad potential of our skill space model, with a demonstrated application to illuminate job evolution directions (Fig.S6).

# Results

### Low-Skilled Occupations Experience the Largest Skill Upgrades

Across the analyzed 727 occupations, low-skilled occupations experience more skill change than high-skilled occupations during 2010-2018; no matter how we distinguish between low and high skill occupations—whether by the average number of core skills across years (Fig.2a), the annual median pay for the occupation (Fig.2b), or the average education requirement (Fig.2c). In other words, re-skilling demand is more significant for workers in occupations requiring low skill complexity and low education with low compensation. The only exception is that jobs requiring master's or doctorate degrees change more than those requiring bachelor's degrees, but both change markedly less than jobs requiring only an associate or high school degree. In Fig.2e, we sorted twenty-two 2-digit SOC occupation categories by decreasing skill change to obtain the face validity of the distinction between low and high-skill occupations. Occupations related to farming, fishing, and forestry, construction and extraction, and transportation and materials moving change most; while occupations related to management, business and financial operations, and computer and mathematical change least. In Supplementary Information Table S4, we present OLS regression models of occupational skill change on skill number, education level, and pay level, showing that the effect of skill level is substantial and consistent. We also show that this pattern still holds when average occupation-level employer concentration, within-occupation job role homogeneity level, and within-occupation job role compositional change are controlled, or when O*NET job zones, which reflect not only formal schooling but also informal experience and training, are used to measure job skill level (see Supplementary Information section 3.1 and 3.6).

Do low-skilled job contents shift in an upskilling direction that suggest potential higher training costs as well as better job quality and prospects for workers? We show in Supplementary Information Fig.S8 that the substantial shift in skills observed within low-skilled occupations signifies an upskilling trend rather than a deskilling one. The skill gap between high- and low-skilled occupations has narrowed from 2010 to 2018, indicating marked transformations in low-skilled occupations that have aligned them more closely with their high-skilled counterparts. This finding is consistent with the trend of rapid earnings growth among low-wage workers that have offset wage inequality in the last decade [24].

**Outsized Skill Upgrades Required in Small Businesses and Labor Markets**

Insofar as skill upgrading is essential, particularly for low-skilled jobs, we now turn to explore a crucial question regarding the organizational and ecological environment of skill change: do jobs at smaller or larger employers and local labor markets face greater pressure for upskilling? On one hand, research suggests that job roles in large firms and markets may change faster. These entities are often leading their fields at the skill frontier and possess the resources to attract and retain highly skilled workers[25–31]. These attributes may require large firms and markets to engage in continuous learning that results in a faster pace of change in job skill requirements compared with smaller companies and markets. On the other hand, smaller firms and markets could pursue more significant skill upgrading to catch up with their larger counterparts. Theories of ecological inertia[32,33] and skill premia[34] also suggest that job roles in large companies and markets may change more slowly and gradually. As companies grow, they develop increased structural complexity that constrains radical change[35]. Moreover, with greater size and density of complementary skills, tasks, and occupations in large organizations and markets, ecological resilience may arise from the scale of network dependencies. This process may increase the ability of large companies and markets to manage uncertainty from their environment[32,33] and reduce the pressure for rapid shifts in job tasks[36–39].

To evaluate these dueling hypotheses, our study focuses on two aspects: first, how the same occupation evolves within small versus large companies and markets over an eight-year observation period between 2010 and 2018; second, development of the skill gap between large and small employers and markets for the same occupations during this period. For the company-level analysis, we classify large employers as those with more than ten annual job postings in both starting and ending years of our analysis, placing them in the top 10% and distinguishing them from 90% of smaller employers. We computed two vector representations for the same occupation for each year, 2010-2018, based on the different skill requirements specified in large and small employers' job posts. This approach captures variations in core skill requirements from both employer size and time. With these vectors, we calculated the level of skill change for a given group of employers (e.g., large employers) as the average distance between their 2010 and 2018 occupation vectors. Additionally, we measured the skill demand gap between the large and small employers in a given year (e.g., 2010) as the average distance between the vectors representing the same occupations but constructed from large and small employers' job ads, respectively.

For the market-level analysis, we categorized job postings into large and small markets. We define local labor markets as commuting zones using 2010 data constructed by Penn State University based on the method developed by the Economic Research Service (ERS) of the U.S. Department of Agriculture[40,41]. Mapping the longitude and latitude locations of job posts to commuting zones, we classify a commuting zone as a large market if its annual job postings ranked within the top 10% among all 652 commuting zones in 2010 and 2018; otherwise, we considered it a small market. As in the firm-level analysis, we construct occupational skill vectors for large and small markets at different timepoints separately based on their different skill requirements for the same occupation in 2010 and 2018. With the same approach as for the small and large company comparison, we computed the average occupational skill change for large and small markets, as well as average occupational skill distance between large and small markets in 2010 and 2018, with very similar findings.

Occupational skill change and the occupational skill gap between small and large companies and markets reveals a strong "skill convergence" effect. Occupations in larger companies and markets exhibit less skill change compared to their smaller counterparts over time (Fig. 3a, c) and the skill gap between small and large companies and markets is narrowing (Fig. 3b, d). This finding suggests that between 2010 and 2018, a significant portion of skill change involved upgrading skills in smaller firms and markets with initially lower skill demands, resulting in small companies and markets undergoing more change and progressively resembling their larger counterparts. This pattern supports the "catching-up" hypothesis that skill change is more granular at the frontier, where most high-skilled occupations, large employers, and markets are located. In contrast, as low-skilled occupations, smaller employers, and markets catch up, they must make more significant advances in skill levels to close the gap with their higher-skilled counterparts. In Fig. 3e, we graph variation in average occupational skill change on a U.S. map for large and small markets, highlighting that upskilling pressures are notably higher in rural regions compared to densely populated coastal states and urban areas. Table S5 and S6 in the Supplementary Information demonstrates that the more significant skill change in small markets and employers is robust after controlling for employer concentration and within-occupation job role compositional change in occupation fixed-effect regressions.

### Who Faces the Most Upskilling Demand?

To understand the implication of skill change on workers, we associate occupational skill change with worker demographics, leveraging statistics from the Bureau of Labor Statistics (BLS) based on the 2018 Current Population Survey (CPS). We measure the reskilling pressure faced by a social group as the weighted sum of occupational skill changes across all occupations, where the weights correspond to the proportion of the group's workers employed in each occupation relative to the group's total employment across occupations. In Fig. 3f, we present Kernel Density Estimation (KDE) curves illustrating each group's employment distribution across occupations of varying skill levels. Compared to female workers, a larger share of male workers are employed in high-skilled occupations requiring a college education or higher. A greater proportion of Latinx and Black workers, relative to their total employment, are employed in low-skilled occupations requiring less than a college education, compared to White and Asian workers. As low-skilled occupations face the greatest reskilling pressure, social groups concentrated in these occupations are more vulnerable to these changes. In Fig. 3g, we compare weighted occupational skill changes between female and male workers and across racial groups. The results show that female workers and non-white workers experience higher reskilling pressure than male and white workers.

# Discussion

Our work advances the empirical science of skills and job change by developing a skill embedding model—skill2vec—that represents occupational skill content and reveals hidden and significant up-skilling demands. Using large-scale job advertisement data coupled with manifold learning methods from machine learning and artificial intelligence[22,42,43], we show that low-skilled occupations experience the greatest skill change when accounting for the distance between skills. This finding directly challenges recent high-profile work[20] that concluded STEM and technology-intensive occupations face the most skill content change from 2007 to 2019.

This result underscores the value of modeling skill change through a continuous geometric representation. Our embedding distance model captures skill proximity through direct and indirect skill overlaps in jobs across the economy. Because jobs are held by skilled persons, and education and training is tuned to help persons obtain jobs, we conceptualize this skill distance in terms of the amount of cognitive and embodied expertise that must be altered for persons to shift from effectiveness in one job to another. In this way, skill distance reflects the training required to perform new skills and jobs. By introducing continuous geometry to supplant cluster and network topology, our model both captures skill interdependencies and generates vectors that permit the efficient and dependable calculation of distance between skills and jobs. Our continuous model allows for better quantification of re-skilling demand than discrete models at the occupational level[20] and enables us to explore variations in skill change for workers across markets, employers, and demographic groups, providing a more nuanced understanding of the human capital system's complex structure.

From this skill space representation, we find that low-skill, less educated, and minority workers who work in small companies and labor markets face the greatest upskilling demand from their jobs. They do not necessarily need to learn more new skills than their high-skilled counterparts, but the new skills they must acquire to retain employment are much more different from their initial skill set, and potentially require more time, expense, and effort in retraining.

We further find a convergence pattern in skill demands. The skill convergence effect we document shows occupations in larger companies and markets exhibit less skill change compared to smaller counterparts over time, with the skill gap between small and large entities progressively narrowing. This finding supports a "catching-up" hypothesis in which skill change is more granular at the frontier, where most high-skilled occupations, large employers, and markets are concentrated. In contrast, low-skilled occupations, small employers and markets require larger leaps in skill space as they catch up. This pattern aligns with theories of ecological inertia[32,33] and skill premia[34] suggesting that job roles in large companies and markets may change more slowly and gradually due to increased structural complexity that constrains it[35].

Our model also reveals the directional flow of skill change. By mapping occupational vectors, we observe nuanced movements at the human–machine interface (see Supplementary Information section 2.7, Fig. S6). For example, card dealers added data skills, while power plant operators adopted environmental auditing. Given the variation in these directional shifts, systematic study is left to future work. We find technological change remains a powerful driver of skill transformation. IT-related skills increasingly diffuse across occupations. Prior work shows that machine-learning relevance predicts skill change[47]. While early skill-biased technological change theory suggested job loss for low-skill workers, recent evidence and our findings show many of these workers instead upgrade their jobs[57,58], demonstrating that workers in routine jobs upgrade their occupations rather than becoming unemployed[56].

Other macroeconomic forces also shape skill demand. Business cycles influence employer expectations. Recessions lead to increased demand for credentials, while tight markets lower entry requirements[59–61]. In Supplementary Information section 3.5 and Fig. S11, we show evidence consistent with prior research that upskilling slowed during periods of low unemployment.

Our findings build on and extend prior approaches to measuring skill distance. Much recent work seeks to build a deeper empirical science of skills and job change through modeling skill distances. The expanding literature includes studies that either use labor flows to infer skill relatedness[44,45] or derive vectors from skill groups to calculate skill distance[46,47]. Nevertheless, such studies consider skills as largely independent from one another. To overcome this limitation, some have used dimensionality reduction techniques to encode skill relatedness with broad, orthogonal categories[48–51], while others have constructed skill networks to model skill interdependencies[21,38,52,53]. But factor analyses do not capture the relative distances between skill categories or between skills within the same category. Similarly, networks do not yield straightforward or continuous distances between skills or jobs, as many more and less paths may connect them.

Several limitations of our approach should be noted. First, the extent of necessary retraining depends on the individual worker's aptitude and preparation. Our data capture employer demand, not the actual cost borne by workers. In Supplementary Section 3.4, we explore this issue by estimating the education costs of re-skilling. Future work should incorporate data on learning costs—time, money, stress—to measure re-skilling pressure more directly.

Second, job mobility and geographic constraints also shape re-skilling needs. Workers may avoid upskilling by relocating, though moves incur costs—transportation, housing, and lost social support. Our analysis does not capture these trade-offs, but they remain important for future work[54].

Third, our data capture shifts in hiring expectations, not current job content. Future research should use on-the-job data to more precisely identify pressure on current workers. Additionally, online job posting data may oversample certain types of positions and underrepresent others, though our validation exercises post strong correlations with official labor statistics.

Despite these limitations, our results raise important implications for low-wage workers. As these workers upskill, they may perform their jobs more productively, which could lead to better compensation—especially if they gain bargaining power. Indeed, the wage gap between low- and median-wage jobs has narrowed in recent years[24]. Wage premiums at large over small firms have also declined[55]. Additionally, skill upgrades may turn "dead-end" jobs into "stepping-stone" jobs that facilitate upward mobility[56]. Future work should examine how skill upgrades shape both compensation and career trajectories.

Our work identifies systemic inequities in the need for re-skilling. Workers in small firms, rural markets, and historically marginalized groups face disproportionate demands to adapt. If met successfully, these pressures may offer opportunities for mobility and improved job quality—but only with appropriate support. Policy responses must therefore attend to regional and organizational variation in how jobs evolve.

As teamwork increasingly dominates the workplace, future research can expand the unit of analysis beyond occupation for skill change and technology-skill complementarities. Large firms and markets—once pioneers of change—may now experience slower, more incremental evolution, while small firms and markets undergo sharper transformations. Future research should investigate whether scale always leads to inertia—or whether periods of surge and catch-up alternate over time.

In sum, our methodological innovation of measuring skill change through embedding distances fundamentally reframes our understanding of who bears the burden of economic transformation and highlights the urgent need for policies that address these hidden inequalities in the evolving labor market.

# Methods

### Data

We analyzed a dataset of more than 167 million job ads from 2010 to 2018 provided by Lightcast. Using a proprietary classification system based on natural-language processing technology and other AI tools, along with dedicated in-house experts, Lightcast. extracted multiple variables from each job ad, including job title, 6-digit Standard Occupational Classification (SOC), location of latitude and longitude values, skill requirements, education requirement, salary, employer name, and more. Lightcast collects information from more than 40,000 job boards and company websites to create the largest dataset of the U.S. labor market[16]. Admittedly, not all new jobs appear online, but online recruitment represents an increasing share of labor market search, even for jobs historically associated with informal and offline recruitment. A 2013 study estimated 60-70% jobs were posted online[62]. Recent research suggested 85%[63]. To verify the representativeness of Lightcast data on the U.S. job market, we calculated occupational demand, pay level, and education requirements using Lightcast data and found that these values are highly correlated with BLS statistics in 2010 and 2018 (see Supplementary Information section 1), justifying the overall consistency and credibility of Lightcast data during the time period of our analysis, despite coverage limitations[16]. Prior work based on Lightcast data have also conducted various validation exercises that compare Lightcast data with the Job Opening and Labor Turnover Survey (JOLTS), CPS, Occupational Employment Statistics (OES), American Community Survey (ACS), and O*NET in terms of SOC occupation level job demand, employment, education, experience, and skill requirements[16,64].

**Training Skill Vectors**

We apply the skip-gram word2vec model[22] to obtain the skill vector representations of jobs. Word2vec learns vector representation of words within a large-scale corpus such that two words frequently occurring in the same direct or indirect linguistic contexts remain close to one another within the latent space, often measured by cosine distance of the angle separating them along the surface[22] and suggesting semantic and/or syntactic correspondence. Using the 167 million job posts describing a set of required skills as instances of training context, we obtain a 200-dimensional vector for Lightcast's 15,182 unique skills. This embedding space encodes the similarity and distance between skills inferred from their direct and indirect co-presence across jobs. A direct co-occurrence between two skills manifests when they appear in a job advertisement together. A first-order indirect co-occurrence between two skills manifests when a third skill co-occurs with the first two skills, creating an indirect skill co-occurrence path. Further indirectness occurs when additional skills are required to connect the two skills along more distant pathways of co-occurrence. This association of directness and indirectness is akin to the direct and indirect pathways through which spectral algorithms underlying PageRank[65] and other Eigenvalue-based centrality scores[66] account for direct centrality, where other nodes connect to a focal node, and indirect centrality, where other nodes connect to nodes that themselves connect with the focal node. Direct co-occurrence disproportionally influences our assessments as the resulting neural embedding proximity approximates the integration of all co-occurrence pathways between skills, and direct co-occurrence represents a much more probable pathway between two skills than indirect co-occurrence.

**Quantifying the Magnitude of Occupational Skill Change**

We calculate the vector of a 6-digit SOC occupation $o$ at a given year $t$ by averaging the vectors of $m$ core skills required by $o$ in $t$. To derive reliable estimates, we focus on 727 active occupations with 100 or more job ads every year and analyze their 5% most frequently required skills in each year, referred to as "core skills". In the Supplementary Information section 3.7, we demonstrate that our findings presented in the main text are robust to alternative definitions of skill content, including expansion of core skills to include all skills and a consideration of skill frequencies as weights. As core skills change over time, the same occupation may manifest evolving vector representations across years even though each skill has only a fixed, globally trained vector. In this way, we focus on occupational skill content change caused by the substitution of core skills[22]. Specifically, we use one minus the Cosine distance between vectors of the same occupation in 2010 and 2018 as the skill change score. We calculate job skill change for all 727 occupations between 2010 and 2018. Calculated scores vary from 0.002 to 0.340, with a median of 0.027. As occupation vectors encode skill distance, this measurement of skill change reflects more precise efforts needed for re-skilling than prior work[20], as illustrated in Fig.1.

Note that the re-skilling cost inferred from an occupational skill change relative to that same occupation at a previous time point is a second-order measurement. This is fundamentally different from important first-order metrics that capture the absolute training cost required to qualify for an occupation from the position of no relevant skill, such as an occupation's: educational requirement, O*NET job zone, or skill number. Our main empirical analysis examines how second-order re-skilling costs vary among occupations with different first-order skill levels (Fig. 2).

Finally, we explored an alternative measurement to characterize occupational skill change as its most significant distance of skill transition in Supplementary Information section 3.2. This approach aims to correct the potential underestimation of skill change for high-skilled vs. low-skilled jobs under the scenario where a high- and low-skilled occupation both add the same number of new skills in 2018 that are extremely dissimilar to their 2010 skills, yet the high-skilled occupation also adds many more similar skills that end up "diluting" the average change associated with the addition of dissimilar skills. With this alternative measurement, we examine individual skill transitions and focus on the largest skill transitions to rule out this possible bias. With this measurement, our conclusion—that lower-skilled jobs are more susceptible to change—remained consistent (see Table S1).

**Validating Skill and Occupation Vectors**

In the Supplementary Information section 2.4-2.6, we present several validation analyses for the skill and occupation vectors. The *t*-SNE visualization of the skill and occupation vectors (Fig. S3-4) show that skills and occupations belonging to the same broader category form meaningful clusters (see Supplementary Information section 2.4). Fig. S5 further demonstrates that occupation skill vectors are consistently closer to their corresponding occupation title word vectors than to other random occupation word vectors.

As the skill embedding space trained on job posting data may skew towards overrepresented occupations and skills in online job space, we validate our embedding space with a large language model pre-trained on a broader corpus of text data from recent work [67]. *Labor Space* is derived from fine-tuning Google's BERT with representative descriptions of different levels of labor market entities from various corpora. In Supplementary Information section 2.5, we show that the relative distances between skills and between occupations in our skill2vec space are consistent with the distances encoded in the LABERT space.

Additionally, we validate the relationship between skill distance by analyzing its ability to predict worker mobility across occupations. To do this, we extracted data on the number of transitions between 1,956 pairs of 6-digit SOC occupations in 2018 from CPS data[68], where individuals reported their occupation from the current and previous year. To assess the skill similarity between each pair of occupations, we calculated the cosine similarity using occupational skill vectors from our 2018 dataset. The Pearson correlation coefficient between skill similarity and logged worker mobility is 0.283 ($p$<0.001), suggesting that when workers shift jobs, they require similar skills. Second, we regress worker mobility against a baseline prediction based on occupation popularity (i.e., the logarithm of the product of employment in the two occupations using 2018 BLS data.) This was done before and after accounting for skill vector distance in the regression. Our findings indicate that including skill similarity significantly enhances the baseline model, with the $R$-squared value increasing by 130% from 0.06 to 0.14.

We also compare the skill embedding approach to the more traditional factor analysis approach used in labor economics work to derive occupational distance measures [49,51]. In Supplementary Information section 2.6, we demonstrate that the distances between occupations, as measured by skill factors, strongly align with those measured using skill embedding vectors. The two approaches also perform similarly in predicting worker mobility with skill2vec explaining slightly more variance in occupation switches.

**Data availability Statement**

The data used in this study were obtained from LightCast under a licensing agreement that does not permit public distribution. Researchers interested in accessing the same data may negotiate access directly with LightCast. Information about LightCast's data offerings and access procedures can be found at https://lightcast.io/.

To ensure transparency and reproducibility, we provide a detailed description of the dataset, including variable definitions and sources, in the manuscript and supplementary materials. We provide a comprehensive variable dictionary for Lightcast data and data processing scripts we used to generate intermediate datasets from raw data at https://github.com/di-Tong/SkillPaper/tree/master/Codes. We also provide all intermediate datasets needed to replicate the main and supplementary results, including aggregated data derived from job postings (such as occupation-year level skill demands and skill embeddings), at https://github.com/di-Tong/SkillPaper/tree/master/IntermediateData. Additionally, the repository includes publicly available datasets used in the analysis: the 2010 Penn State University Labor-Sheds for Regional Analysis data [https://sites.psu.edu/psucz/data/], 2018 CPS data [https://cps.ipums.org/cps/], O*NET job zone data [https://www.onetonline.org/], 2010 and 2018 BLS Occupational Employment Statistics (OES) data [https://www.bls.gov/oes/tables.htm], and Labor Force Statistics data [https://www.bls.gov/cps/cps_aa2018.htm].

For further inquiries regarding data access, researchers may contact LightCast directly or reach out to the corresponding author for guidance on the data request process.

**Code Availability**

The python code used in the analysis is available at https://github.com/di-Tong/SkillPaper/tree/master/Codes.

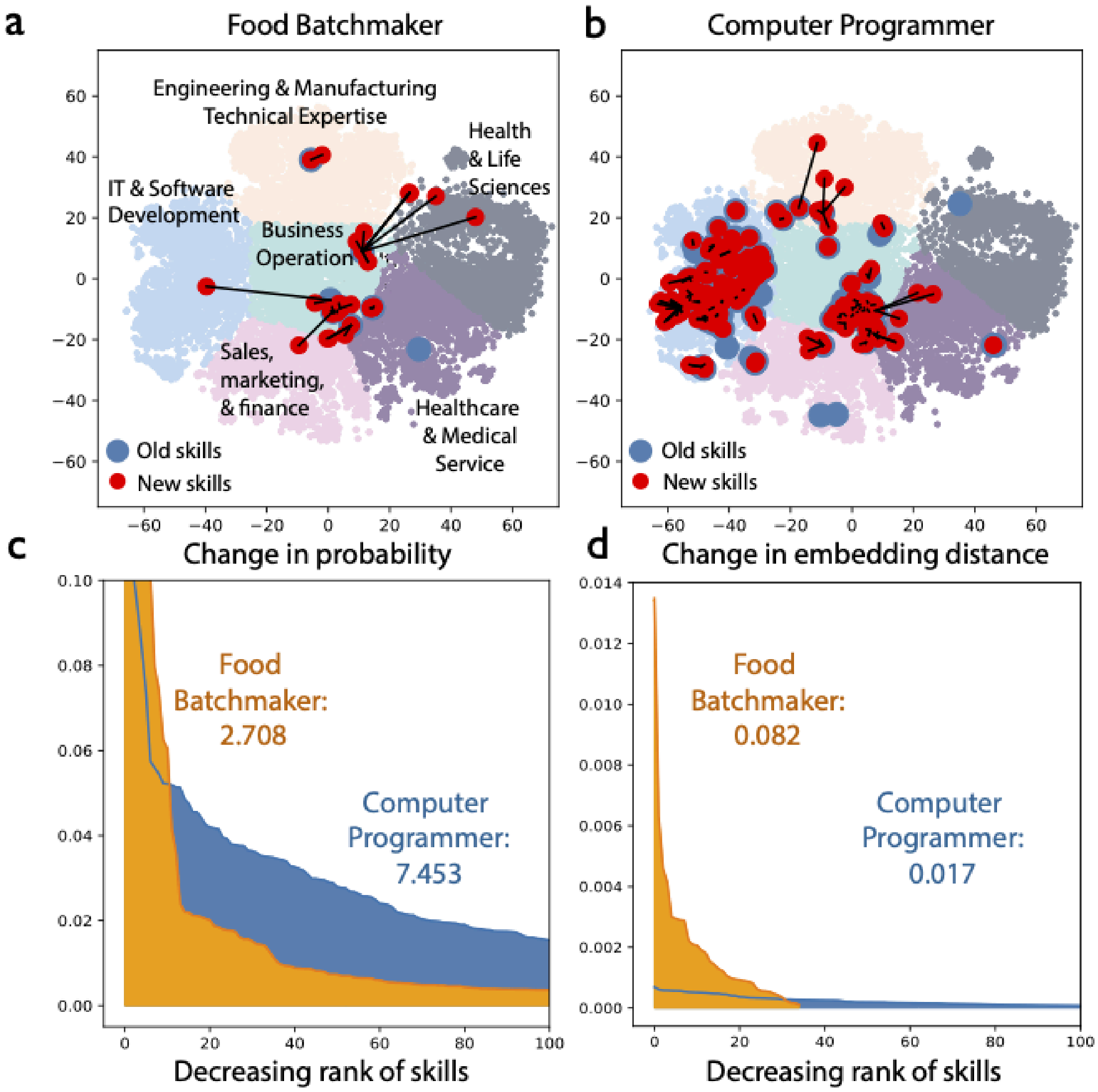

**Fig. 1: Measuring occupational skill content change without considering skill proximity induces biases. a**, Food Batchmakers have fewer new 2018 core skills (with top 5% occurrence) than computer programmers, but the new skills are on average located much further from 2010 core skills. We illustrate the proximity and distance between skills by representing them in 6 well-formed clusters detected with the K-means algorithm from the t-SNE 2-dimensional skill vectors transformed from the original 200-dimensional skill embedding. Each dot represents a 2-dimensional skill vector. Each new core skill (red dot) of food batchmakers is linked to its nearest 2010 old skill (blue dot), with the length of link proportional to the distance associated with skill transition. Compared with panel b, there are many fewer red dots in panel a, but the distance between the red dots and their nearest blue dots are typically larger. **b**, Computer programmers have many new core skills in 2018 not among their 2010 core skills, but 2018 and 2010 skills are very similar. **c**, Without considering skill distance, computer programmers are assessed as experiencing substantially higher skill change than food batchmakers, 2010 to 2018. Each unit of the $x$ axis corresponds to a skill ranked from highest to lowest in terms of skill probability change. The $y$ axis denotes skill probability change for each skill (see Supplementary Information section 2.2 for method details). The area under the orange and blue curves sum up the job-level skill change for food batchmaker and computer programmer, respectively (sums are reported in the annotation text within the plot). **d**, After controlling for skill distance, food batchmakers are associated with much larger skill change than programmers. Each unit of the $x$-axis corresponds to a core skill ranked from highest to lowest in terms of their contribution to the job-level skill embedding distance change. The $y$-axis denotes skill embedding distance change attributed to each skill (see Supplementary Information section 2.2 for method details).

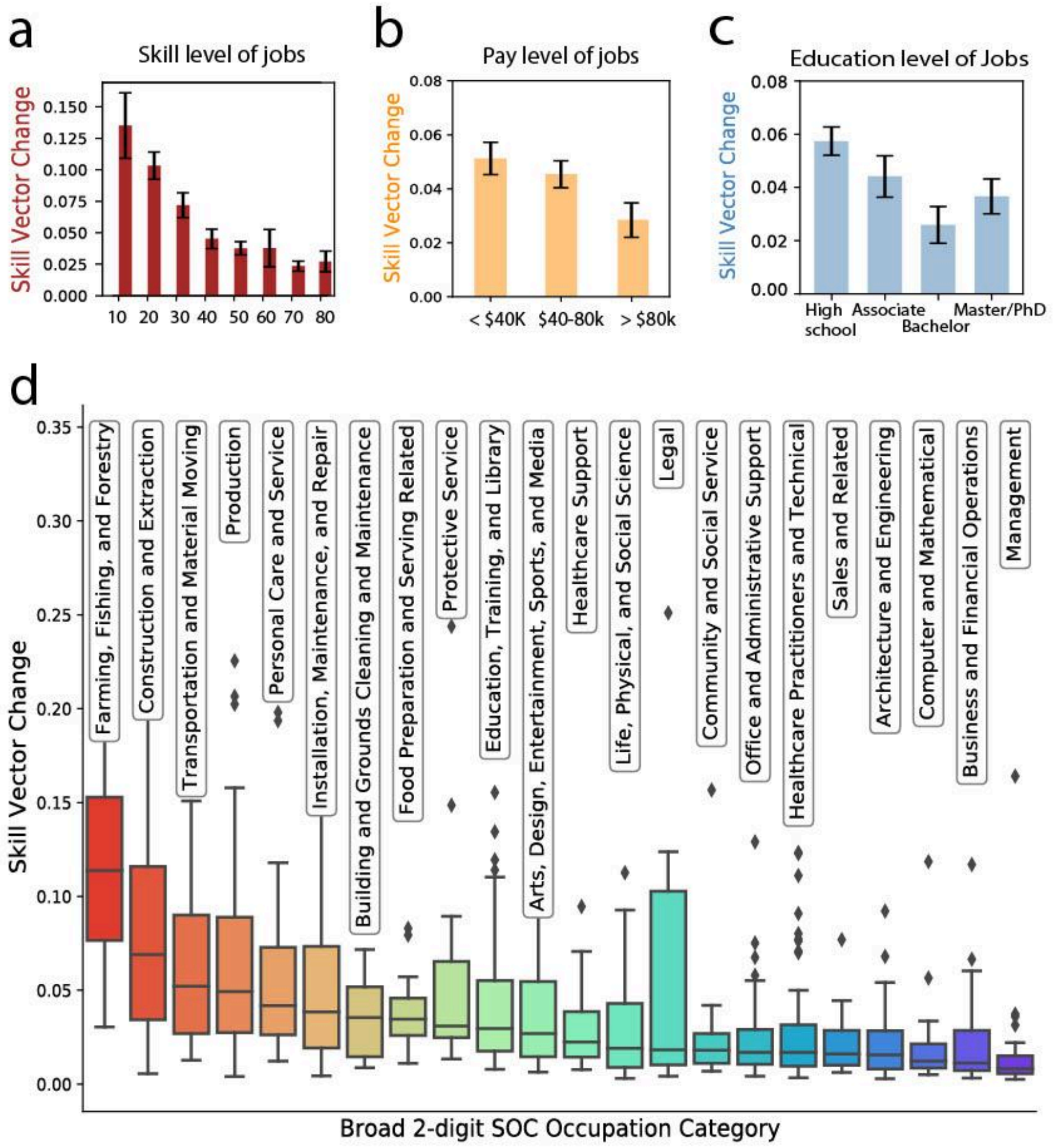

**Fig. 2: Low-skilled occupations experience the largest skill change during 2010-2018 when skill distance is accounted for. a**, Occupations with lower skill complexity have higher average skill content change. Each bar denotes the average skill vector change for a group of jobs whose core skill number falls within the range labeled on the *x*-axis. **b**, Occupations with lower pay have higher average skill vector change. Each bar denotes the average skill vector change for a group of jobs whose average annual median pay falls within the range labeled on the *x*-axis. **c**, Except for graduate degrees, occupations with lower education requirements change more; jobs requiring masters or doctorate degrees change more than those requiring bachelor degrees. Each bar denotes the average skill vector change for a group of jobs whose average entry education requirement falls within the range labeled on the *x*-axis. **d**, The skill vector change distribution for 6-digit SOC occupations among each 2-digit SOC occupation category. Bins are ordered from the group with the largest median change to the one with the lowest median change.

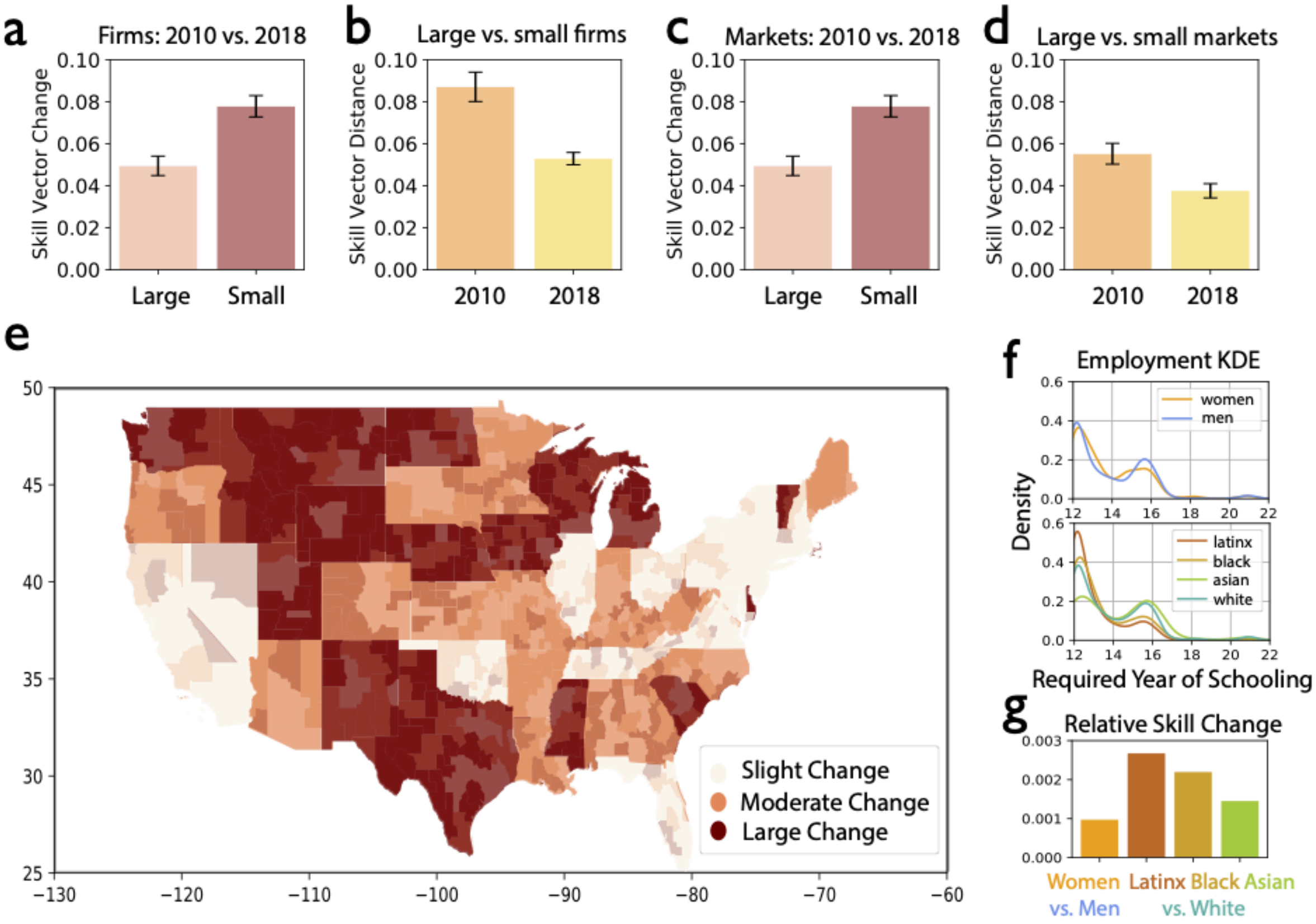

**Fig. 3: Variation in skill demand change for different organizations, local labor markets, and social groups. a,** Small firms have a higher level of occupational skill change than large firms, 2010 to 2018. Bars denote average skill vector change for the same group of occupations in large and small firms. **b,** Small firms upgrade skill requirements and converge to large firm skill demands, 2010 to 2018. Bars denote average occupational skill vector distance between large and small firms, 2010 and 2018. **c,** Small markets have a higher level of occupational skill change than large firms, 2010 to 2018. **d,** Small markets upgrade skill requirements and converge to large firm skill demands, 2010 to 2018. **e,** Mapping occupational skill change in U.S. regions. States and commuting zones are grouped into three quantiles in terms of their average occupational skill change from low to high, labeled as "slight", "moderate", and "large change". **f,** A larger proportion of male workers, relative to their total, are employed in high-skilled occupations compared to female workers; A larger proportion of latinx and black workers, relative to their total, are employed in low-skilled occupations compared to white and asian workers. The Kernel Density Estimation (KDE) curves show the distribution of employment across occupations with varying required education years for six social groups. **g**, Female workers and non-white workers face higher-reskilling pressure than male workers and white workers. The first bar represents the difference between female and male workers' weighted occupational skill change. The other three bars denote the difference in weighted occupational skill change between three minority groups and white workers.

**Acknowledgments**
We thank Lawrence Katz and workshop participants at MIT Institute for Work and Employment Research for helpful comments. We also thank Bledi Taska and the staff at Lightcast for generously sharing their data and comments. J. E. thanks the NSF SBE-1829366, AFOSR FA9550-19-1-0354 and DARPA HR00111820006 for support, and L. W. acknowledges the support of Richard King Mellon Foundation and NSF SOS:DCI-2239418.

# Supplementary Information for

## **Low-skilled occupations face the highest up-skilling pressure**

Di Tong, Lingfei Wu, James A. Evans

Correspondence to: jevans@uchicago.edu (J.A.E.)

## CONTENTS

## 1. Supplementary Data: Representativeness of Job Ads Data Collected by Lightcast

Lightcast data could be biased for multiple reasons. They may include duplicate job ads and oversample high skill jobs which are more likely than low skill jobs to appear in online posts. Moreover, among the analyzed job ads, only 50% specified educational requirements and 17% listed salary. To address these concerns, we have compared Lightcast data against the 2010 and 2018 Occupational Employment Statistics (OES) assembled by U.S. Bureau of Labor Statistics (BLS) in job demand / employment, education requirement, and salary for the entire sample of occupations. We confirm that these two data sources are highly consistent in all three variables. Specifically, we find the Pearson correlation coefficient $r$~0.8 ($p$<0.001) for labor market share (Fig. S1a), $r$~0.8 ($p$<0.001) for salary (Fig. S1b), and $r$~0.9 ($p$<0.001) for education (Fig. S1c).

In Fig. S1a, each dot represents a 6-digit SOC occupation. 786 Occupations can be matched between Lightcast job post data and BLS-OES data in 2018; 759 occupations can be matched between the two datasets in 2010. The $X$ axis denotes the log value of Lightcast vacancy post number for each occupation. The $Y$ axis denotes the log value of BLS estimated employment for each occupation. The Pearson correlation between the log of Lightcast occupational vacancy post number and the log of BLS occupational employment is 0.8 in 2018 and 0.75 in 2010. The magnitude difference between BLS occupational employment data and Lightcast occupational demand data could be attributed to the fact that employment includes incumbent workers not reflected in labor demands, and that not all jobs hire workers through online job ads and a single job post may seek to hire more than one worker.

Fig. S1b presents 772 occupations that can be matched between Lightcast and BLS-OES datasets in 2018; and 743 occupations that can be matched between the two datasets in 2010. $X$ and $Y$ axes denote the log value of Lightcast and BLS average annual median salary for each occupation, respectively. The Pearson correlation between the log of Lightcast and BLS occupational annual median pay is 0.87 in 2018 and 0.83 in 2010.

In Fig. S1c, each dot represents an occupation in 2018, summing to 682 matches between Lightcast entry education requirement data and BLS education and training assignments by detailed occupation data in 2018. The $X$ axis denotes the average Lightcast entry education year requirement for each occupation in 2018. The $Y$ axis denotes the BLS estimated typical entry education year for each occupation in 2018. Here 12 refers to High school diploma or equivalent; 14 to Associate's degree; 16 to Bachelor's degree; 18 to Master’s degree; 21 to Doctoral or professional degree. Note that in Lightcast’s job posts, education year could only take a value from the set {12, 14, 16, 18, 21}, denoting different degrees {“high school”, “associates”, “batchelors”, “masters”, “doctorate”}. Nevertheless, because we calculated the average education year from all job posts for each occupation, the occupational average education year could take on value between those in the set. BLS, on the other hand, only provides entry degree level for each occupation: “No formal educational credential”; “High school diploma or equivalent”; “Some college”, “No degree”; “Postsecondary nondegree award”; “Associate's degree”; “Bachelor's degree”; “Master's degree”; “Doctoral or professional degree”. In order to compare the two datasets, occupations with “No formal educational credential” are left out; “Some college”, “No degree” as well as “Postsecondary nondegree award” are coded as 12 (the same as “High school diploma or equivalent”). The Pearson correlation between Lightcast and BLS 2018 occupational entry education level is 0.92.

We also verify the Two-step Deduplication Process that Lightcast used[63], in which the key component is to build advanced parsing engine to extract and normalize a number of data elements from each job listing, including job title, job ID, source, posting date, employer name, location, job description text, etc., and then use these variables to screen for duplicates. In our data analysis, we did not find job posts that duplicated all of these fields.

## 2. Supplementary Methods

### 2.1 Formula and Replication of Deming and Noray’s (2020) Job Skill Change Measurement

As is demonstrated in equation 1, Deming and Noray (2020)[20] measure the skill content change of occupation *o* as the sum of the absolute value of difference in share for each skill from 2007 to 2019, in which a given skill $s$’s share of occupation *o* in year *t* is defined as the proportion of *o*’s job ads that require $s$ in year *t*. In equation (1)a, $\left(\frac{skill_o^s}{JobAds_o}\right)_{t1}$ is the number of job ads of occupation $o$ in year *t1* that require skill $s$ divided by the number of job ads of occupation $o$ in that year. Similarly, $\left(\frac{skill_o^s}{JobAds_o}\right)_{t0}$ is this share of skill $s$ for occupation $o$ at *t0*. The absolute value of the difference between these *t0* and *t1* skill shares is calculated for all skills required by occupation *o* in either *t1* or *t0*. The sum of these absolute skill share differences, $SkillChange_o$, captures the skill change for occupation *o* from *t0* to *t1*.

To account for a secular increase in job post number and skill number per post,[20] Deming and Noray weight the skill change rate calculated from equation (1)a by multiplying the inverse growth rate of the average number of skills per job in an occupation. Specifically, this is measured as the ratio of skill occurrence divided by post number in 2007 to that in 2019, for each occupation (see equation (1)b). In equation (1)b, the numerator is the number of skills required for occupation *o* in *t0* divided by its required skill number in *t1*. The denominator is the number of job posts of occupation *o* in *t0* divided by its number of job posts in *t1*. This weight is not a correction against a principled null model, but simply weights occupations that grow in complexity less heavily than those that do not.

$$SkillChange_o = \sum_{s=1}^{S} \left\{ Abs\left[ \left( \frac{skill_o^s}{JobAds_o} \right)_{t1} - \left( \frac{skill_o^s}{JobAds_o} \right)_{t0} \right] \right\} \quad (1)a$$

$$Weight_o = \frac{\frac{NSkill_{o,t0}}{NSkill_{o,t1}}}{\frac{NAds_{o,\,t0}}{NAds_{o,\,t1}}} \quad (1)b$$

We replicate this measurement on the same job ads dataset compiled by LightCast, limiting the sample to posts with non-missing employer and MSA information. Our replication for 6-digit SOC occupation skill change rates from 2007 to 2019 highly correlates with the scores listed in [20] appendix with 0.87 Pearson correlation. The replicated measure does not obtain exactly the same values for two reasons: (1) We only have access to the first 5 month of job posts data in 2019, whereas [20] uses 10 months' data for 2019; (2) [20] further filter the sample by only retaining posts with employers that could be matched to a specific Compustat dataset inaccessible to us.

**2.2 Occupational Skill Change Measurements in Figure 1c and d Illustration**

In Fig. 1c, the area under the curve of each occupation corresponds to Deming and Noray's (2020)[20] job skill change measurement discussed in section 2.1. Note that all skills appearing in the job ads of a given occupation are taken into account. Because the measurement sums up the probability change for each skill required by the job in either start or ending year, we can easily decompose the total skill change for an occupation and rank skills by the proportion of occupation change for which they account. Fig. 1c presents the individual skill change for computer programer and food batchmaker in a ranked order from highest to the lowest.

For Fig. 1d, we first calculate the occupational skill vector change for Food Batchmaker and Computer Programmer using the method described in Methods section and applied to our main analyses. Namely, we represent the occupation skill content in 2010 and 2018 with the average vector of top 5% core skills required in those two years, respectively. With these occupation vector representations, we measure occupational skill change as one minus the cosine distance between occupation vectors in 2010 and 2018.

We then locate two groups of skills that contribute to the occupation skill change and attribute occupation level change to each skill. For each new core skill that emerges in 2018, we remove it and recalculate the occupation vector representation in 2018 based on the remaining skills. We then use the cosine distance between this adjusted 2018 occupation representation and the 2010 representation to estimate the occupational skill vector change rate that would have occurred if a given new skill were not added in 2018. We approximate the change that a given skill accounts for as the absolute value of the difference between the real occupation skill vector change rate and this controlled occupation skill vector change rate. Similarly, for each removed core skill that only appears in 2010, we estimate the adjusted occupation skill vector change rate that would have arisen if this skill were absent in 2010. Using the same approach, we then estimate the change contributed by each removed skill.

In Fig.1d, for each occupation, we rank all these new 2018 core skills and removed 2010 core skills by their estimated individual contribution to the occupation level change in skill embedding distance. With a comparable arrangement with Fig. 1c, the area under the curve for each occupation in Fig. 1d sums up these individual skill level contributions and approximates occupation level skill distance change. Note that because our measure for occupational skill change does not naturally decompose to individual skill level as Deming and Noray's (2020)[20] approach, the area under the curve estimated from Fig. 1d (0.066 for food batchmaker, 0.027 for computer programmer) is slightly different from the real occupational-level skill change calculated by our holistic approach and applied for all analyses (0.082 for food batchmaker, 0.017 for computer programmer). Therefore, we report the real occupational level change in Fig. 1d text annotation to stay consistent with the rest of the paper.

### 2.3 Cluster Approach to Measure Occupational Skill Change

In order to construct a more conservative test of our conclusion regarding the reskilling burden for low-paid and low-educated occupations relative to that of Deming and Noray[20], we develop a version of occupational skill formally equivalent to Deming and Noray's approach, but using data-driven skill clusters identified from the skill co-occurrence network rather than individual skills.

We used 2010 Lightcast job postings data to construct a skill co-occurrence network, a topological representation of skills linked by co-presence within job advertisements[21]. Our approach adds geometric precision to create a Pointwise Mutual Information (PMI) skill network. We first calculate the PMI of each pair of skills using equation (2). With each skill represented as a node in the network, an edge between two nodes is added if the PMI between the two skills is larger than 0, implying that these two skills are more likely to co-occur in the same or similar job posts than expected under independence[69]. The weight for each edge is the PMI score between the two skill nodes connected by the given edge. The 6 skill communities detected from the PMI network by the commonly-used Louvain community detection algorithm are: (a) business skills; (b) engineering, technical, and physical skills; (c) programming skills; (d) clerical and administrative skills; (e) scientific knowledge; (f) health and medical care skills. The modularity of the partition is 0.51, indicating that this division is very strong, and represents substantial community structure in the network[70].

For skill *i* and *j*,

$$PMI_{i,j} = \log \frac{p_{i,j}}{p_i \times p_j} \tag{2}$$

where *p* refers to probability and

$$p_{i,j} = \sum_{v=1}^{V} p(i|v) \times p(j|v) \times p(v) \tag{3}$$

denotes the probability of skill $i$ and $j$ co-occuring in the same or a similar job post $v$.

Utilizing the 6 communities automatically detected from the PMI network, we calculated skill community change with equation (4). Fig. S2a shows that this measurement reverses the assessment of Deming and Noray[20]. Fig. S2b suggests that this measurement highly correlates with our main skill vector change measurement.

$$SkillChange_o = \sum_{sc=1}^{SC} \left| \left( \frac{NSkill_o^{sc}}{NSkill_o} \right)_{t_1} - \left( \frac{NSkill_o^{sc}}{NSkill_o} \right)_{t_0} \right| \tag{4}$$

**2.4 Embedding Validation: Internal Validation**

We created *t*-SNE visualization for the vectors of 6652 skills that have non-missing skill family labels classified by the data vendor Lightcast (Fig. S3). We color code the skills with skill family labels. Similarly, Fig. S4 is the t-SNE visualization for 6-digit SOC occupation vectors, color-coded with 2-digit SOC occupation group names. In both figures, skills and occupations belonging to the same broader group form meaningful clusters. Similar skill and occupation groups are also close to each other in the space. These visualizations demonstrate the effectiveness of the skill2vec space in capturing the relative distances between skills.

Furthermore, we compared occupational skill vectors with corresponding occupation title word vectors derived from the skill2vec space to assess if they are consistently closer to each other than with other random word vectors. As the skill2vec space consists of 15,182 skills in the Lightcast job postings data, many words in occupation titles do not exist in this corpus. Therefore, we focus on 132 occupations with parts of its title represented in the skill embedding space and calculate their corresponding title vectors as the average of the matched phrases vectors. As a baseline comparison, we generate an average random word vector for each occupation by randomly selecting the same number of phrases as the matched occupation title phrases from the skill embedding space. For example, two phrases associated with the occupation title Public Relations and Fundraising Managers are represented in the skill embedding space: public relations and fundraising. The title vector of this occupation is the average vector of these two phrases. And the random word vector is the average of two randomly selected phrases from the skill embedding space. We calculated the cosine similarities between average occupational skill vectors and their corresponding title vectors, then compared these values to the cosine similarities between average occupational skill vectors and random word vectors.

Fig. S5a plots the distribution of these two groups of cosine similarities for the 132 occupations. The former group is generally much larger than the latter random baseline group. A two-sample *t*-Test shows that the mean of the similarity between occupation skill and title vectors is statistically significantly larger than the mean of the similarity between occupation skill and random word vectors ($t$-statistic=30.248, $p$<0.001). Fig. S5b is a scatter plot where the $x$-axis denotes the cosine similarity between an occupation's average skill vectors and average title vectors, whereas the $y$-axis represents the cosine similarity between the average skill vectors and average random word vectors. All occupation dots are below the diagonal line benchmarking the equivalence of the two values, confirming that occupational skill vectors are consistently much closer to their corresponding occupation title word vectors than with random word vectors derived from the same embedding space.

**2.5 Embedding Validation: Pre-trained LLM**

As *skill2vec* is trained solely on job postings, it may skew towards overrepresented occupations and skills in the online job space. To address this potential bias, we validate our embedding space using a large language model pre-trained on a broader corpus of text data: *Labor Space*[67]. *Labor Space* is derived from Google's BERT, which is trained on Wikipedia and the Google Books Corpus. Kim et al.[67] fine tuned BERT with representative descriptions of different levels of labor market entities from various corpora, including Occupational Information Network (O*NET) and European Skills, Competences, Qualifications, and Occupations (ESCO). *Labor Space* therefore captures the semantic distances between labor market's key elements.

We focus on 259 occupations and 7,219 skills in our data, as each token within these has an identical match in the *Labor Space*. We calculated the pairwise cosine similarity between all occupation pairs and skill pairs using the *Labor Space* word embedding representation and our *skill2vec* skill embedding representation, respectively. We then compared the two sets of pairwise cosine similarities obtained from the two spaces. For occupation pairs, the Pearson correlation between the two sets of pairwise cosine similarities is 0.34 ($p \sim 0.00$). For skill pairs, the Pearson correlation between the two sets of pairwise cosine similarities is 0.29 ($p \sim 0.00$). We also generated a random baseline reference by randomly reshuffling occupation and skill names in LABERT vectors before calculating correlations. The random baseline correlations are 0.00 ($p = 0.89$) for occupation pairs and 0.00 ($p = 0.36$) for skill pairs. This analysis suggests that the relative distances between skills and occupations in the skill2vec space are consistent with that encoded in the LABERT space.

**2.6 Embedding Validation: Factor Analysis**

In this section, we compare the skill embedding approach to the more traditional factor analysis approach used in labor economics work to derive occupational distance measures[49,51]. To identify latent skill factors, we use 2018 job postings data to construct an occupation-skill probability matrix. Each occupation-skill cell consists of the number of job posts for the occupation requiring the given skill divided by the number of all job posts for the occupation. To generate a larger training dataset while maintaining model runtime efficiency, each occupation is weighted by either 0.1% of its total job post number or 1, whichever is greater. This weighting approach results in a dataset of 28,068 rows, providing a larger sample size than previously used for identifying skill factors in other works. To prepare for factor analysis, we first remove extremely low-variance skill columns and skills that are perfectly or almost perfectly correlated with other skills in the matrix. This preprocessing reduces skill variables from around 15000 to 1554. We then standardize the matrix to have a mean of 0 and a standard deviation of 1.

Next, we apply factor analysis to identify 200 orthogonal skill factors. Together, these factors account for 88% of the variance in the occupational skill probability matrix. We transform the occupational skill probability matrix with the factor loadings to represent each occupation as a vector of 200 factor scores. With these 200-dimensional occupation factor vectors, we calculate the distances between all pairs of occupations and compare them with the pairwise vector distances calculated from skill embedding representation of occupations. The Euclidean distance or cosine similarity between occupation pairs represented by factors post moderate, statistically significant correlations with the distances calculated based on skill embedding representation of occupations: the pearson correlation is 0.52 for Euclidean distance and 0.59 for cosine similarity (both $p < 0.001$).

The two approaches also perform similarly in predicting job switches based on 2018 CPS data, with skill2vec explaining slightly more variance in job switches. We regress worker moves between a pair of occupations on the pairwise occupation distances and a baseline prediction based on occupation popularity (i.e., the logarithm of the product of employment in the two occupations using 2018 BLS data). We found that skill2vec-based distances explain more variance in job transitions than factor-based distances, regardless of how distance is defined. With factor-based occupation-pair distances, the $R$-squared in the regression is 0.095 for Euclidean distance, and 0.137 for cosine similarity. When pairwise occupation distances are measured with skill embedding approach, the $R$-squared in the regression is 0.131 (37% increase) for Euclidean distance, and 0.141 (3% increase) for cosines similarity.

### 2.7 Identifying the Direction of Occupational Skill Change with Skill Atoms

Occupation vectors predict not only the magnitude of skill change but also its direction. First, we construct a "coordinate system" of the skill space by using the discourse atom topic modeling approach, which performs $k$-SVD matrix factorization on skill vectors to accurately and efficiently label the skill space[71,72]. The derived vectors or "skill atoms"[71] represent near-orthogonal axes capturing the essential "bases" of distinct human capacity, which can be linearly combined to recover the vector representations of 15,182 actual skills. Specifically, each skill is represented as a linear combination of $k$ skill atoms. We trained models by setting the atom number $k$ from 50 to 500. The model performs best with 210 atoms based on a balance between (1) $R^2$, which measures how well the atoms predict all skill vectors; and (2) topic diversity, which measures how distinct the atoms are from one another.

After we obtain the 210 skill atoms to anchor our skill-space as coordinates, we specify the direction of occupational skill change within this system. We quantify the rise and fall of skill atoms as a function of how all 727 occupations shift collectively. Rising atoms are those a majority of occupations approach, and declining atoms are those a majority of occupations depart. Specifically, we calculate the overall importance of a skill atom as the sum of its weights across all occupations in that year and compare how overall importance changed between 2010 and 2018 following the procedures detailed below.

1) The compositionality of occupations based on skill atoms:
   a) Map each occupation to 5% core skills filtered by skill probability
   b) Map each skill to atoms - Denote each skill $s$ with its weights on each atom $j$:

$$s = \sum_{j=1}^{210} weight_{sj} \tag{5}$$

Note that according to the current model, there are only 5 non-zero $weight_{sj}$ for each skill $s$.

c) Map each occupation to atoms - Denote each of the 727 occupations in a given year (2010 or 2018) as a combination of atoms by adding up its core skills represented by atoms in the equation (6). The weight of each atom for each occupation is normalized by dividing the sum of all atom weights for the given occupation. For an occupation $L_t$ with $S_t$ core skills at time point $t$:

$$L_t = \sum_{j=1}^{210} weight_{otj} \tag{6}$$

in which

$$weight_{otj} = \frac{\sum_{s=1}^{s_t} weight_{sj}}{\sum_{j=1}^{210} \sum_{s=1}^{s_t} weight_{sj}} \tag{7}$$

2) The overall importance change of skill atoms on the job space:

a) Measure the overall importance level for each skill atom on the job space at a given time point - for each atom $j$ at each time point $t$, summing up its normalized weight for each occupation $o$ ($weight_{otj}$):

$$importance_{jt} = \sum_{o=1}^{727} weight_{otj} \tag{8}$$

b) Measure atom overall importance change on the job space between 2010 and 2018 - for atom $j$:

$$ImportanceChange_j = importance_{j_{2018}} - importance_{j_{2010}} \tag{9}$$

Fig.S6a presents these 210 skill atoms in a matrix of 14 rows and 15 columns. To demonstrate the relative location of all 210 atoms on a 2-D graph, we first apply the T-SNE transformation on the original skill atom vectors to reduce their dimension from 200 to 2. We then construct a grid of 15 columns and 14 rows on the area between the lowest and highest values for each of the 2 dimensions for all atoms. Finally, for each node in the grid, we assign the nearest unassigned atom to occupy it. Finally, we employed two human coders to label these 210 skill atoms "human"' or "machine"-related based on the closest 25 skills to a given skill atom in the space to observe how these two kinds of skill atoms rose or declined in the past decade. Fig.S6b shows two exemplary occupations and their most dramatically altered atoms, 2010 to 2018. We present skill atoms that decline in importance across all jobs and atoms that increase in importance in this period (Fig.S6c-d) to highlight global transformations of skill in the U.S. labor market. Future work can build on these spaces to explore other causes and consequences of the rise and fall of skill atoms for skill change across distinct occupations.

## 3. Supplementary Results

### 3.1 Job Zone and Skill Change

We use O*NET five-level job zone classification (i.e., "little or no preparation", "some preparation", "medium preparation," "considerable preparation", "extensive preparation") as a proxy for occupational skill level. The job zone measure (https://www.onetonline.org/help/online/zones) reflects on-the-job training and experience in addition to formal education in capturing the first-order learning costs associated with each job—their relative difference from no education. Fig. S7 demonstrates that skill change decreases as job zones increase from 1 to 5 except that jobs in zone 5 change more than those in zone 4.

### 3.2 Occupational Skill Change as the Most Significant Skill Transition

In the main paper, we measure occupational skill change as the distance between the occupational average skill vector at different time points. This approach may underestimate the skill change of high-skilled occupations compared to low-skilled occupations under a situation when both groups of occupations add the same number of distant new skills, and the high-skilled occupations also add many more similar new skills. With this skill-adding structure, the high-skilled occupations change more than the low-skilled occupations, yet the large number of similar skill additions may end up "diluting" the significant change embodied in the addition of dissimilar skills. To address this potential bias, we developed an alternative measurement to characterize occupational skill change as its most significant distance of skill transition, and tested whether low-skilled occupations still change more than high-skilled counterparts.

For each occupation $o$, we identify the $n$ newly-added core skills in 2018. For each of these new skills, $s_n$, we find its nearest skill (based on embedding distance), $s_{n10}$, among the 2010 core skills of $o$. We consider the embedding distance between $s_n$ and $s_{n10}$ as reflecting the minimum amount of skill transition needed for a worker who held $o$ in 2010 to acquire $s_n$. Calculating such distance from all pairs of $s_n$ and $s_{n10}$, we use the maximum pairwise skill transition distance to measure the most significant individual skill transition a worker who held $o$ in 2010 needs to experience to keep their job in 2018. To rule out the influence of outliers that may arise at the maximum boundary, we also investigated measurements based on various top positions within the distribution of individual skill transition. Specifically, we considered the pairwise skill transition distance that ranks the top 5% and 10% when ordered from the largest to the smallest.

In Table S1, we regress these three measurements of occupational skill change on occupational skill complexity (natural log of average core skill number), natural log of average occupational annual wage, and average education year, respectively. Results show that low-skilled occupations still have larger skill change than high-skilled occupations when skill change is measured as the most significant individual skill transition.

### 3.3 Upskilling of Low-Skilled Jobs

To understand whether the significant skill change that took place in low-skilled jobs denoted an upskilling direction, we calculated the skill gap between the low-skilled job group and high-skilled job group in 2010 and 2018 and examined whether the gap has narrowed over the years. In Fig. S8a-c, we divided occupations into two groups based on three different criteria: (a) occupations with core skill number at or above median core skill number across all occupations are classified into the high-skilled group, and the rest is classified into the low-skilled groups; (b) occupations with average annual pay at or above the median across all occupations are classified into the high-skilled group, and the rest is classified into the low-skilled groups; (c) occupations with college or above education requirement are classified into the high-skilled group, and the rest is classified into the low-skilled groups.

We calculate the skill gap between the high-skilled and low-skilled occupation group in a given year by pairing nearest occupations across the two groups and average the pairwise distance. Specifically, for each occupation in the high-skilled group in 2010, we pair it with its nearest occupation (based on 2010 average skill vector) among all low-skilled group occupations in 2010. Following the same procedure, we find the nearest occupation in the high-skilled group for each low-skilled group occupation in 2010. The skill distance between the high-skilled and low-skilled occupation groups in 2010 is measured as the average of all these pairwise occupation skill vector distances in 2010. The left yellow bars in Fig.S8a-c represent the skill distance between the high-skilled and low-skilled occupation groups in 2010, whereas the right green bars in Fig.S8a-c represent the inter-group skill distance in 2018, calculated based on the same approach described above. The three panels all suggest an 'upskilling' story of low-skilled occupations moving closer to high-skilled occupations in terms of skill requirements from 2010 to 2018.

### 3.4 Meaning and Significance of Occupational Skill Change: Re-Educational Costs

We further explore the meaning and significance of occupational skill change by demonstrating that distance in the skill space correlates with the amount of re-education required for workers to move from one distribution of skills to another. To draw comparison between jobs with marginal skill differences, we undertake the following analysis:

1. We randomly select a 10% sample from all 2010 job posts, and identify among them pairs of job posts, $x$ and $y$, from the same occupation, where post $y$ encompasses all skills listed in post $x$, but also possesses additional skills. Table S2 presents two job ad examples picked from January 2010 data to illustrate the type of comparison we are making in calculating the extra education necessitated by marginal skill change. The first example includes two ads for administrative assistant positions at entry and senior levels. Both ads require basic administrative skills, such as communication and administrative support, as well as proficiency in Microsoft Office tools like Excel, Word, and PowerPoint. The senior position, however, also requires specialized skills in customer management, such as Siebel CRM, Portal Tools, Customer Information Control System (CICS), Direct Mail, and Newsletters. Here, the skill distance between the two ads is 0.17, and the senior role requires an additional 2 years of education compared to the entry-level role (16 vs 14 years). The second example includes two job ads for registered nurses with different levels of specializations. Both positions require basic patient care and service skills, whereas the position that specializes in Interventional Radiology additionally requires more advanced skills on this direction, such as Advanced Cardiac Life Support (ACLS), Catheterization, Catheterization Laboratory (CATH LAB), Critical

Care, and Interventional Radiology (IR). The skill distance between the two job posts is 0.15 and the education year difference is 2 years.

2. Using the sample selected in step 1, we further select pairs where *y* requires more education years than *x*.
3. Using this subsample, we plot the distribution of the number of additional skills in *y* compared with *x* (which typically follows a unimodal distribution, see Fig. S9), and identify the peak of that distribution $m_o$. This peak is more likely than other numbers in the distribution to correspond to a regularly-added combination of skills, typically obtained through more advanced education. Fig. S9 illustrates eight example occupations, each showcasing distinct distribution types peaking at varying numbers of additional skills.
4. Next, we analyze a subset of 6,585,524 job post pairs from the Step 1 sample where *x* and *y* belong to the same occupation *o*, and *y* requires $m_o$, $m_o$-1, or $m_o$+1 more skills beyond the common skill set shared with *x*. Because our data do not allow a comprehensive measurement of all sorts of re-training for skill acquisition and there are only a few levels of education degrees, we apply step 4 to upweight cases where formal education requirements differ between job posts within the same occupation. Note that in the analytical sample built in this step, job post *y* may add $m_o$ skills but does not require a higher degree than *x* because either those skills are not necessarily the regularly-added combination of skills, or could be obtained without a new formal degree. We aim to test whether larger skill distances predict degree requirement changes and more substantial education year differences by comparing a range of cases, encompassing those with no meaningful degree changes, those with such changes, and variations in the magnitude of change between them. In the first column of Table S3, we regress the pairwise education year difference on the embedding distance between the average skill vectors of the jobs. These variables show a positive and statistically significant relationship.
5. Because job posts pairs within each occupation tend to present similar pattern, we aggregate the data to the occupation level by calculating for each the average required education year difference and skill distance based on three groups of job posts pairs: pairs with $m_o$, $m_o$*-1,* or $m_o$*+1* skill additions, respectively, resulting in 1257 data points.

Fig. S10 is based on this aggregated analysis and shows that marginal skill distance predicts educational difference. The 2nd column in Table S3 presents the regression at occupation level and shows a stronger positive correlation between skill distance and education requirement difference. We then transform skill distances into z-scores in the 3rd column of Table S3 to facilitate the interpretation of results. We find that one standard deviation increase in skill distance is associated with 0.172 years more schooling.

To ensure these patterns are not a production of this specific sample, we applied different random selection approaches to draw samples from 2010 data, including with different seeds or randomly select a full month data. Based on the six different random samples we examined, the association between education year difference and skill distance is always positive and statistically significant. The average of the Pearson correlation of these variables is 0.05 with raw job post unit of analysis, and 0.14 with aggregated occupation level of analysis. The average coefficient from regressing education year difference on the z-score of skill distance is 0.1 at raw job post level, and 0.17 at aggregated occupation level. We note that this measured shift in educational *level* does not account for the necessary shift in educational *type*. For example, shifting from a data analyst job in a media company to that in a biotech company may not require a higher level of education (e.g., a Bachelor degree), but is more likely to require a degree with a different major as distance in the skill space increases.

### 3.5 Labor Market Tightness and Skill Change

We computed the year-on-year average occupational skill changes (e.g., 2010-2011, 2011-2012, etc.) across all occupations from 2010-2018 and found a positive correlation with the yearly civilian unemployment rate released by BLS: pearson coefficient 0.95 with p-value less than 0.005. Fig. S11 shows that the yearly average occupational skill changes decline with the yearly unemployment rate from 2010-2018. Given that our skill change measure largely captures upskilling directions, this pattern is consistent with the business cycle literature [59,60,73]. As the labor market becomes tighter from 2010 to 2018, employers have gradually eased off their upskilling attempts in skill requirements as they decrease education and experience requirements.

### 3.6 Robustness Check: Controlling for Employer Concentration, Within-Occupation Job Role Homogeneity, and its Changes

This and the following section present regression-based robustness checks for the two findings that (1) low skilled occupations experience more skill changes than high-skilled occupations; and (2) large firms and markets experience less skill changes than their small counterparts. The regressions used to verify the first finding use occupation as the unit of analysis and weight each occupation by the average number of job posts in 2010 and 2018 to adjust for the fact that some occupation cells are larger and have more accurate averages. To verify the second finding, we ran two sets of regressions with different analytical units. The first one uses the firm-commuting zone(cz)-occupation as an analytical unit to track within firm-region-occupation changes. Each firm-cz-occupation unit is weighted by the corresponding average number of job posts in 2010 and 2018. Because this analysis could only be performed on the subsample of units that appeared in both 2010 and 2018, we ran the second set of regressions to include occupations that disappeared or newly appeared in 2018. In the second set of regressions, the analytical units are firm-cz that appeared in both 2010 and 2018. Each firm-cz is weighted by the corresponding average number of job posts in 2010 and 2018.

The regressions in this section address several potential biases. First, for occupations with high average local employer concentration, their skill change could be overwhelmingly determined by changes driven by a few employers. Therefore, we control for employer concentration in regressions to ensure that the variation in job skill change by job skill complexity, employer size and local market size does not merely reflect variation in occupational employer concentration within the local labor market.

Following[74–76], we measure employer concentration with the Herfindahl-Hirschman Index (HHI) of the share of Lightcast vacancy postings from each employer for each SOC 6-digit occupation within a local labor market in a given year, as specified in equation (10). To align with the geographical unit used in examining job skill variation by local labor market size, here we identify the local labor market as commuting zones.

$$EmpConcentr_{o,k,t} = \sum_{i=1}^{N} \left( \frac{JobAds_{i,o,k,t}}{\sum_{i=1}^{N} JobAds_{i,o,k,t}} \right)^2 \tag{10}$$

where $JobAds_{i,o,k,t}$ refers to the number of Lightcast vacancy advertisements posted by employer $i$ on occupation $o$ in local market $k$ during year $t$. In the occupation unit regressions, we aggregate this measure to occupation level as average occupation employer concentration across all local markets, year 2010 and 2018. In the firm-cz-occupation unit regressions, we take the 2010 and 2018 averages for employer concentration at occupation-cz level. And for firm-cz unit regressions, employer concentration is aggregated to cz level, averaging across all occupations in year 2010 and 2018.

Another potential bias lies in the measurement of occupational skill change: it reflects both shifts in within-job-role skill requirements and job role composition within an occupation. A 6-digit SOC occupation is a meaningful unit of analysis for our purpose because it is a relatively consistent framework for analyzing individuals' roles in the labor market over time. Based on 2018 CPS-ASEC data, approximately 87% of observations (with non-missing occupation variables) stay in the same occupation as the prior year. Workers in the same occupations tend to share similar human capital, job task requirements, working conditions, and career trajectories. However, skill requirements do vary within occupation and across but sometimes also within firms and regions. An occupation can encompass multiple closely connected but not completely identical sub-occupational job roles. When some of the sub-occupational job roles look too different from others, it may raise the concern that they are misclassified. While job role composition change is also an important part of occupational skill change, we want to ensure that it is not driving the results, especially when potentially misclassified job roles blend in.

Therefore, we construct a measure for occupational job role composition change from 2010 to 2018 with changes in within-occupational job role homogeneity level. With each job post represented as the average required skill vectors, an occupation's job role homogeneity is measured by the average pairwise job post similarity for all job posts of the given occupation in a given year. A higher score means less within-occupation variance: a score of 1 means all job posts are identical in skill contents. Changes in this metric reflect whether an occupation becomes more or less diverse in its job role composition by adding or abandoning job roles that are very different from the main job roles. To single out skill changes from job role composition changes, we control for within-occupation job role homogeneity level change in all the regressions. Additionally, occupations internally more homogeneous might be in the more stable job fields, embed in more specific contexts, and therefore become prone to have less dramatic skill changes. Therefore, we also control for occupational job role homogeneity to rule out any potential confounding effect.

For occupation unit regressions, we randomly sampled 5% job posts for each occupation in a given year to construct the within-occupation job role homogeneity measure. The average occupational homogeneity score based on 2010 and 2018 data is 0.63. For firm-cz-occupation unit regressions, we use all job posts for each unit to measure within firm-cz-occupation job role homogeneity. The average firm-cz-occupation homogeneity score based on 2010 and 2018 data is 0.95. For firm-cz unit regressions, we take the average of within firm-cz-occupation job role homogeneity for each firm-cz unit.

Table S4 presents OLS regression results explaining job skill change variation with the three different measurements for job skill level: job skill complexity, pay, and education. Consistent with Fig. 2 in the main body, models 1-3 show that low-skilled occupations change more. Models 4-5 demonstrate that this pattern still holds when employer concentration, within-occupation job role homogeneity level and its temporal change are controlled. The coefficients for employer concentration align with the findings in Hershbein et al[76]—occupations with larger local employer concentration experience more re-skilling.

Table S5 presents a series of occupational fixed effect model results describing how skill content change for the same occupation varies with organization and local market size, based on the sample of firm-cz-occupations that appear in both 2010 and 2018. The occupation fixed effect controls for time-invariant occupational characteristics, including within-occupational job role homogeneity level. Model 1 and 2 show that larger firms and larger local markets experience smaller skill changes for the same occupation than their smaller counterparts. As shown in model 3-5, these results are still robust when employer concentration and job role composition change is taken into consideration. When we put both employer and market size together in the same model with all controls, the coefficient of employer size is a bit smaller with larger standard error ($p$=0.065) compared to when market size is not controlled. This suggests that employer size and market size are correlated and that the latter has a more robust and statistically independent association with the within firm-cz-occupation skill change. Surprisingly, when employer and/or labor market size is controlled, larger employer concentration correlates with lower job skill change. Given that employer concentration is constructed at a relatively narrow region level, this pattern might result from the high collinearity between local market size and employer concentration.

While table S5 focuses on within firm-cz-occupation skill changes, Table S6 examines firm-cz level overall skill changes that incorporate disappearing and newly-appearing occupations in 2018. The weighted firm-cz unit regressions show that larger firms and local markets have smaller overall skill changes. Coefficients for firm size and market size in Table S6 are smaller than their coefficients in Table S5, suggesting that variations in skill changes by entity sizes are slightly less pronounced for overall entity skill changes than entity-occupation skill changes. Such differences could be due to the different samples involved or that larger firms and markets include more rapidly-changing occupations.

**3.7 Robustness Check: Different Job Content Scope with Skill Weights**

While the Lightcast job postings data provide rich and valuable information on dynamic job skill requirements that enables analyses in this paper, the data also create the new task of identifying the relative importance of skills to jobs as compared with the O*NET data. In the main analysis, we adopt a discrete approach to account for skill importance differences in representing occupational skill content—limiting occupational skill composition to the most common 5% skills for a given occupation. This 5% threshold is arbitrary, however, and a discrete approach is not necessarily better than a continuous approach that assigns weights to each skill. Therefore, we present a series of robustness checks here with different job content scopes (all skills, top 50% core skills, and top 25% core skills) using a continuous approach to incorporate skill importance differences: we represent occupation vectors as the sum of weighted skill vectors, as shown in equation (11).

$$Vector_{o,t} = \sum_{s=1}^{S_t} \frac{Freq_{o,s,t}}{Freq_{o,t}} \times Vector_s \tag{11}$$

where the weight is the proportion of occurrences of skill *s* in job posts on occupation *o* in year *t* among the sum of occurrences for all skills that appear in job posts on occupation *o* in year *t*.

Table S7 consists of three sets of estimations of model 4 in the main table S4 that explain job skill change variation with skill complexity using all skills, top 50% core skills, and 25% core skills to define job skill content, respectively. Similarly, Table S8 and 9 present estimations of model 5 from Table S5 and S6, respectively, testing the variation of skill change by organization and local market size with different job content scopes. Note that tables S8 and S9 have more observations than tables S5 and S6 because the main 5% core skill change approach is only applied to units (firm-cz-occupation or firm-cz) with at least 5 core skills in both 2010 and 2018. The current continuous skill change measures loosen the criteria to 5 total skills in both 2010 and 2018 and therefore largely expands the sample, especially after weighting. All findings are robust to the continuous skill weight approach and different thresholds for job skill content definition.

## 4. Supplementary Figures

### 4.1 Figure S1.Validation of Lightcast data representativeness

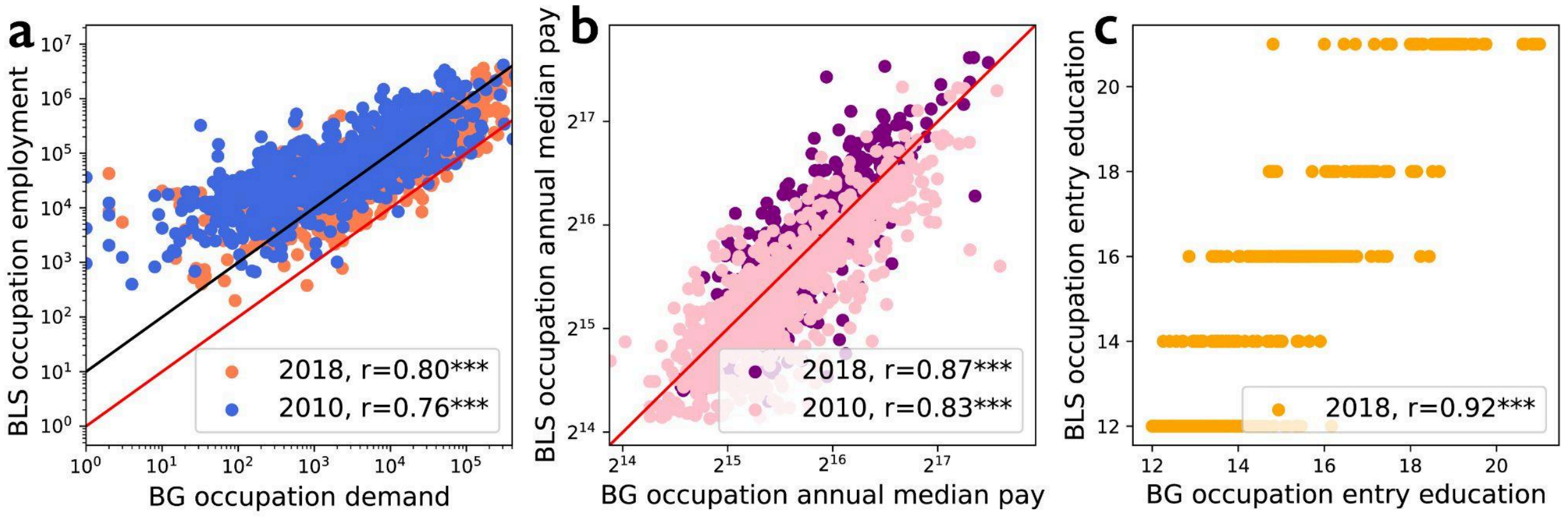

**Supplementary Figure 1. Validation of Lightcast data representativeness: a**, Lightcast occupational demand is highly consistent with the state of occupational employment in the U.S. Coral dots are occupations in 2018, while royal blue dots are occupations in 2010. The red line is the diagonal $y = x$. The black line shows the red line's vertical translation upward. **b,** Lightcast occupational median pay accurately represents occupational median pay in the U.S. Purple dots are occupations in 2018, while pink dots are the same in 2010. The red line is the diagonal $y = x$. **c**, The Lightcast occupational entry education level information could be trusted to represent the occupational entry education requirement in the U.S. Each dot represents an occupation in 2018.

### 4.2 Figure S2. Cluster approach to measure occupational skill change

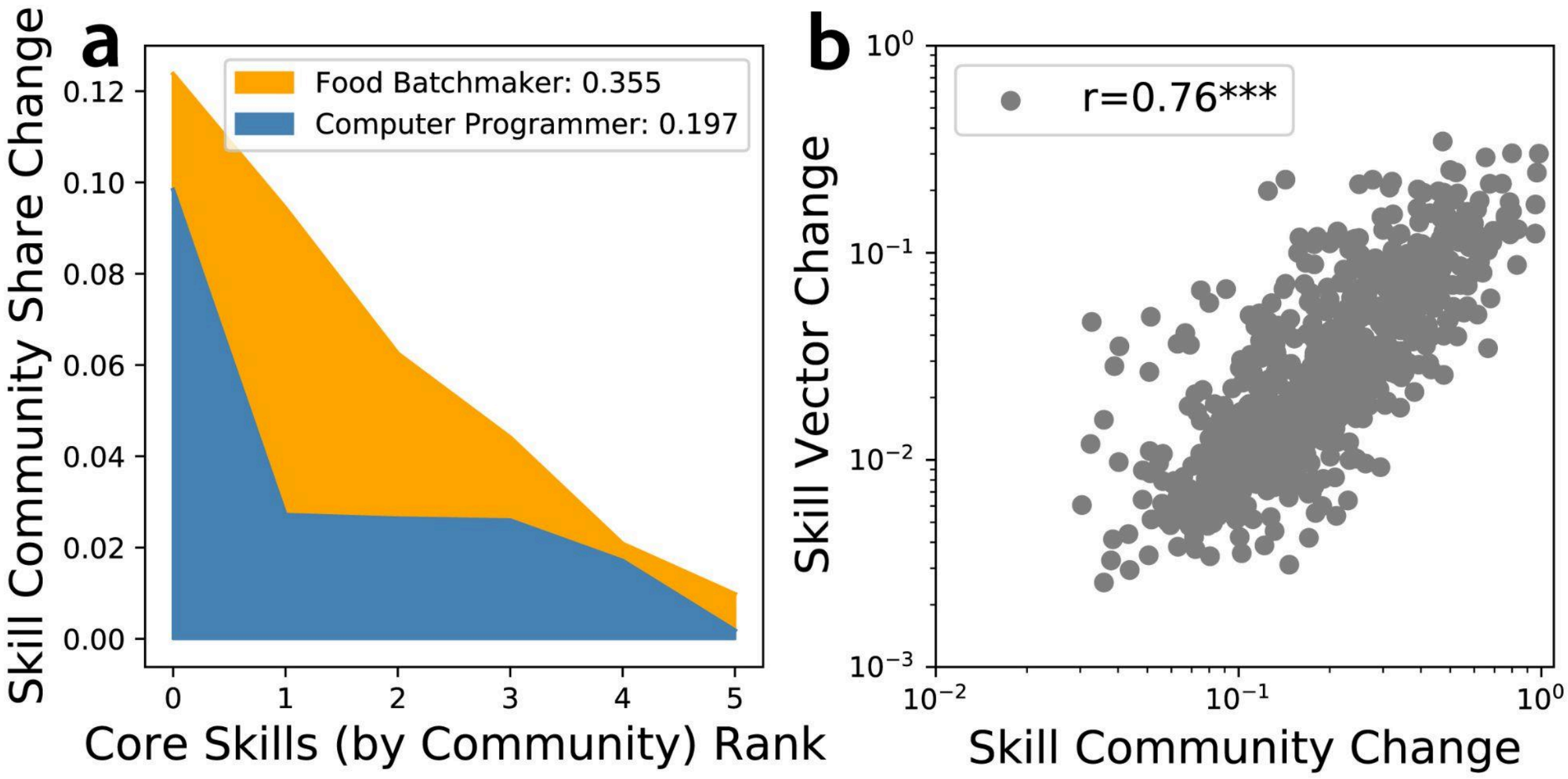

**Supplementary Figure 2. Cluster approach to measure occupational skill change: a**, After roughly controlling for skill distance through tracking skill community share change instead of individual skill probability change, food batchmakers are associated with much larger skill change than programmers. Each unit of the *x* axis corresponds to a skill community ranked from highest to lowest in terms of skill community share change. The *y* axis denotes skill community share change for each skill community. Area under the curve (AUC) of skill community share change for each occupation demonstrates the sum of skill community changes for all 6 skill communities from 2010 to 2018. **b**, Occupation skill change measured by skill community share change and skill vector change highly correlate with a 0.76 Pearson correlation.

## 4.3 Figure S3. T-SNE visualization of skill vectors

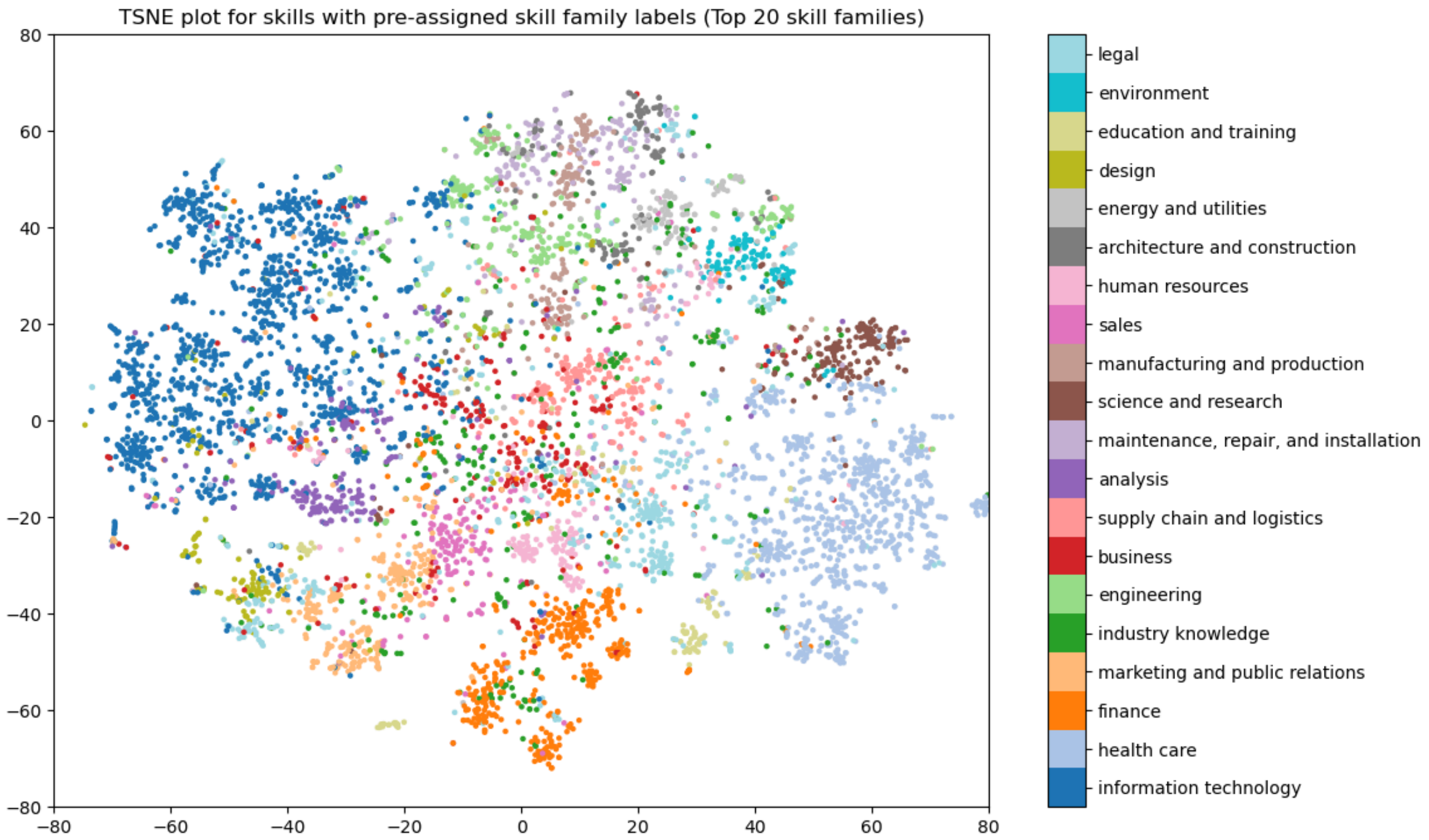

**Supplementary Figure 3. T-SNE visualization of skill vectors:** Each dot corresponds to the t-SNE representation of skill vectors for 6652 skills with 28 skill family labels classified by the data vendor lightcast. The skills are color-coded with skill family labels (only the most frequently appeared 20 labels are presented ).

### 4.4 Figure S4. T-SNE visualization of occupation vectors

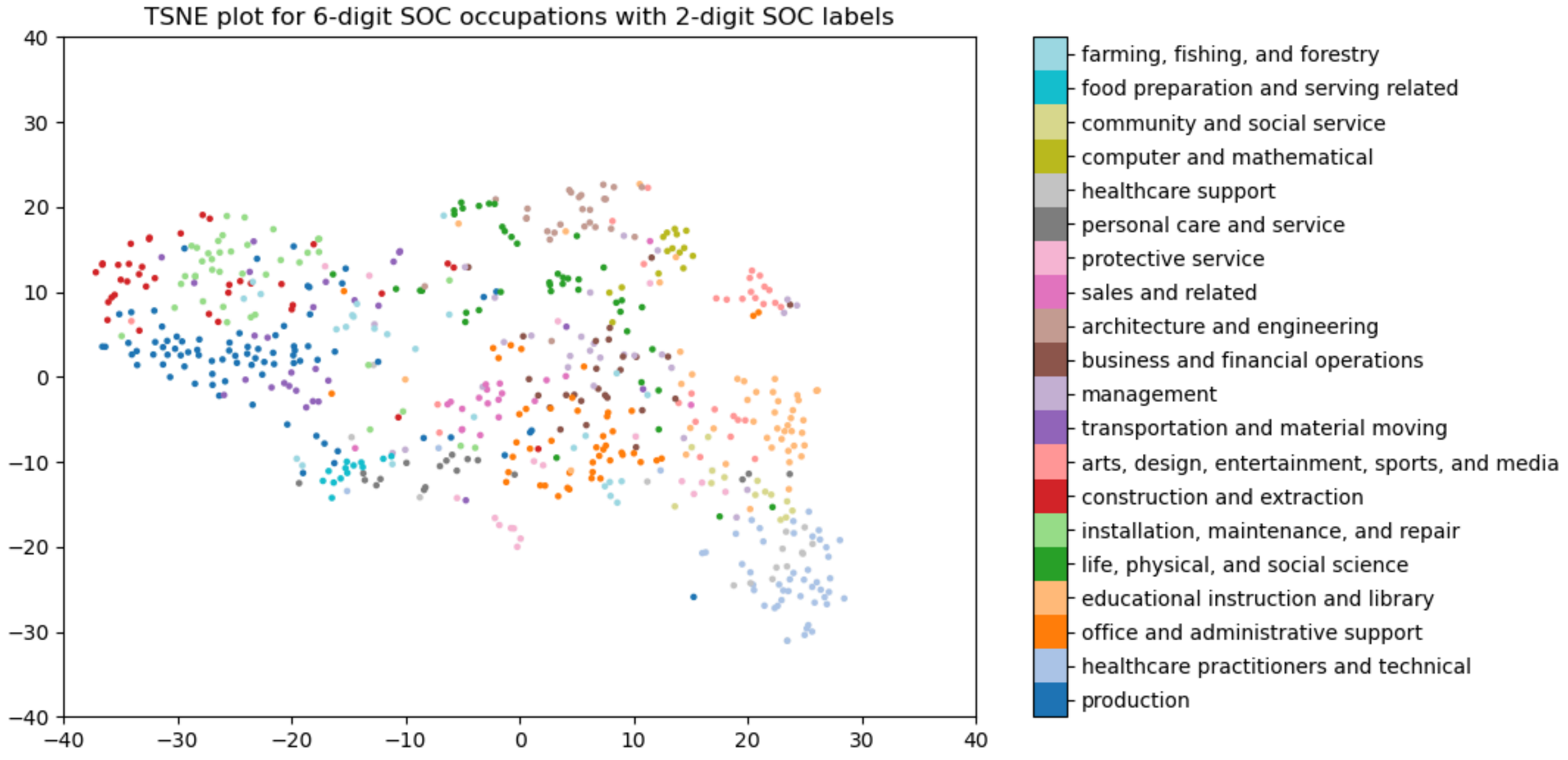

**Supplementary Figure 4. T-SNE visualization of occupation vectors:** Each dot corresponds to the t-SNE representation of 6-digit SOC occupation vectors color-coded with 2-digit SOC occupation group names (only the most frequently appeared 20 groups among the 23 groups are presented ).

### 4.5 Figure S5 Occupation skill vectors and title word vectors comparison

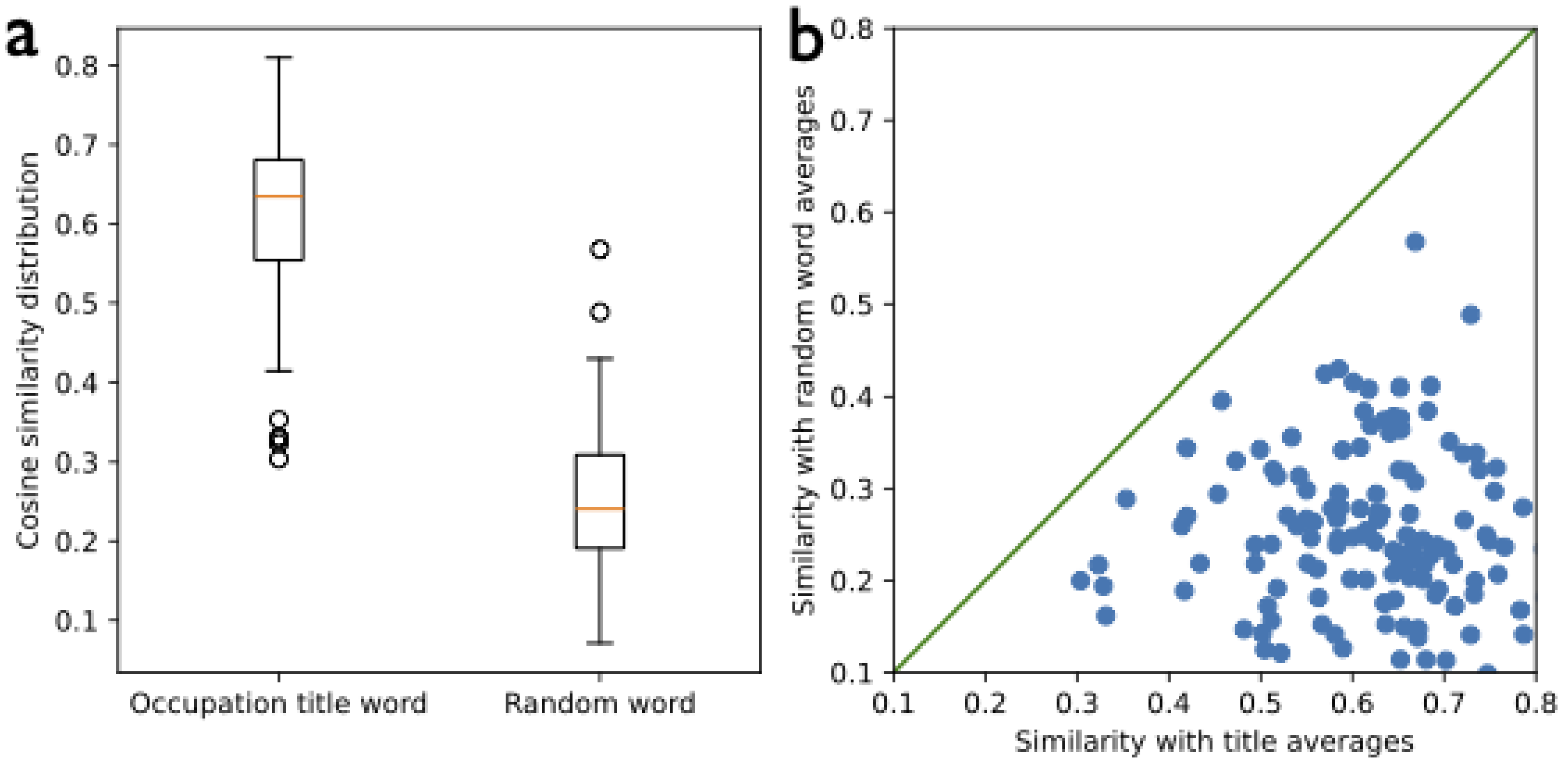

**Supplementary Figure 5. Occupation skill vectors and title word vectors comparison: a,** The box plot on the left presents the distribution of cosine similarities between occupational average skill vectors and corresponding occupation title word vectors for 132 occupations. The box plot on the right shows the distribution of cosine similarities between occupational average skill vectors and random word vectors. **b,** Each dot represents an occupation, with x-axis denoting the cosine similarity between the occupation's average skill vector and its average title word vector, whereas y-axis corresponding to the cosine similarity between the occupation's average skill vector and average random words vector.

### 4.6 Figure S6. Occupations' re-skilling direction illustrated by skill atoms

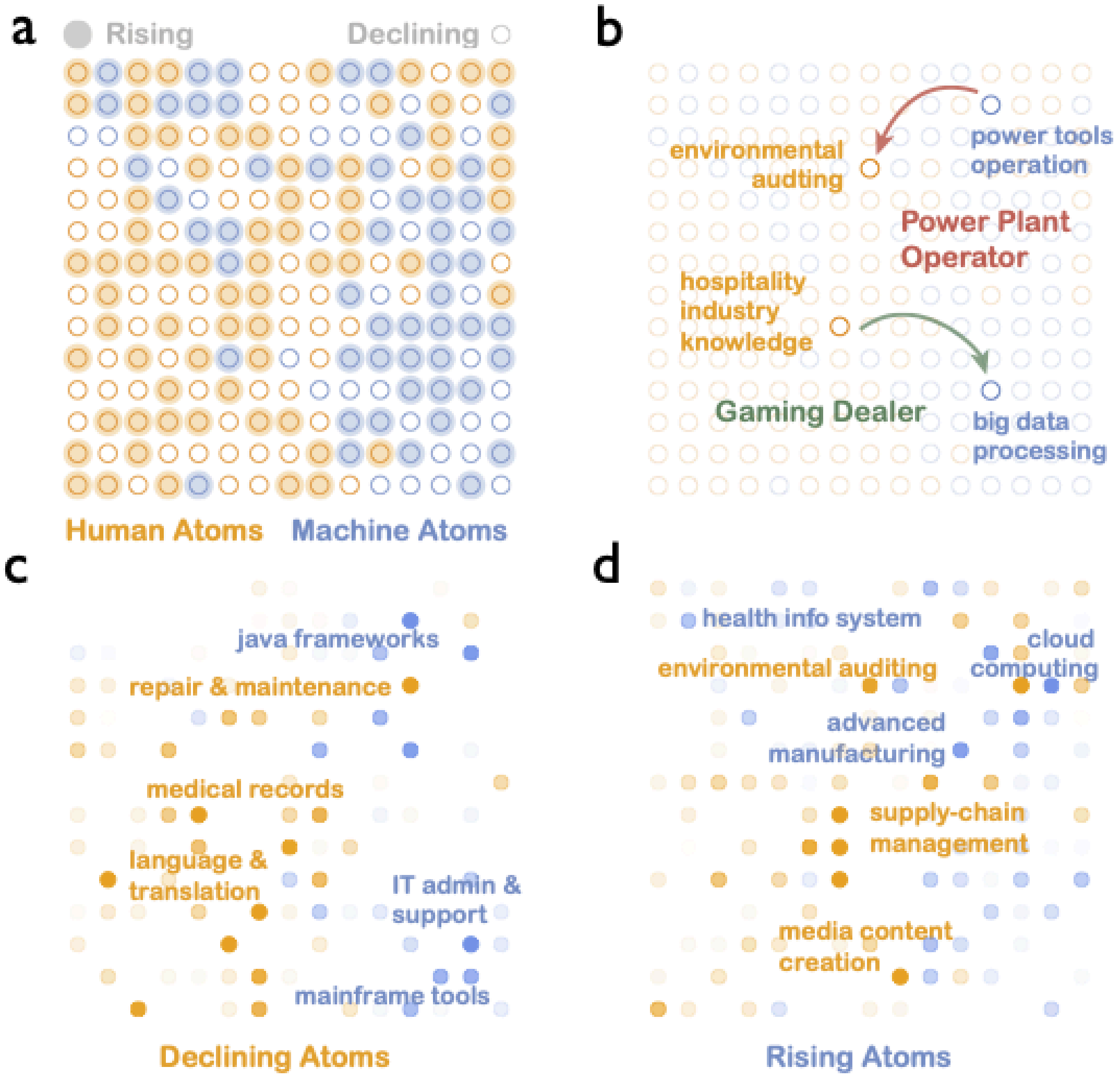

**Supplementary Figure 6: Occupation re-skilling direction illustrated with skill atoms. a**, Each dot denotes a latent skill atom learned from the skill embedding space, embodying a meaningful dimension in the skill space and embedding its relationship to other skill atoms as well as to each specific skill. Orange skill atoms are defined predominantly by the human interface, and blue atoms are defined by requiring machine-operation and/or interface. Filled dots are skill atoms with rising importance on the skill demand space between 2010 and 2018, whereas the empty dots represent declining skill atoms. **b**, Examples of the re-skilling direction for individual occupations. For Power Plant Operators, the power tools operation atom declines the most and the environmental auditing atom rises most in importance (2010-2018). For Gaming Dealers employed by casinos, the skill atom that declines most in importance from 2010 to 2018 is hospitality industry knowledge, whereas the one that increases in importance most is big data processing. **c**, Declining skill atoms with negative (<0) importance change in the job space: the less transparent an atom, the larger its importance declines. **d**, Rising skill atoms with positive (>0) importance increase: the less transparent an atom, the larger its importance increases.

## 4.7 Figure S7. Job zone and skill change

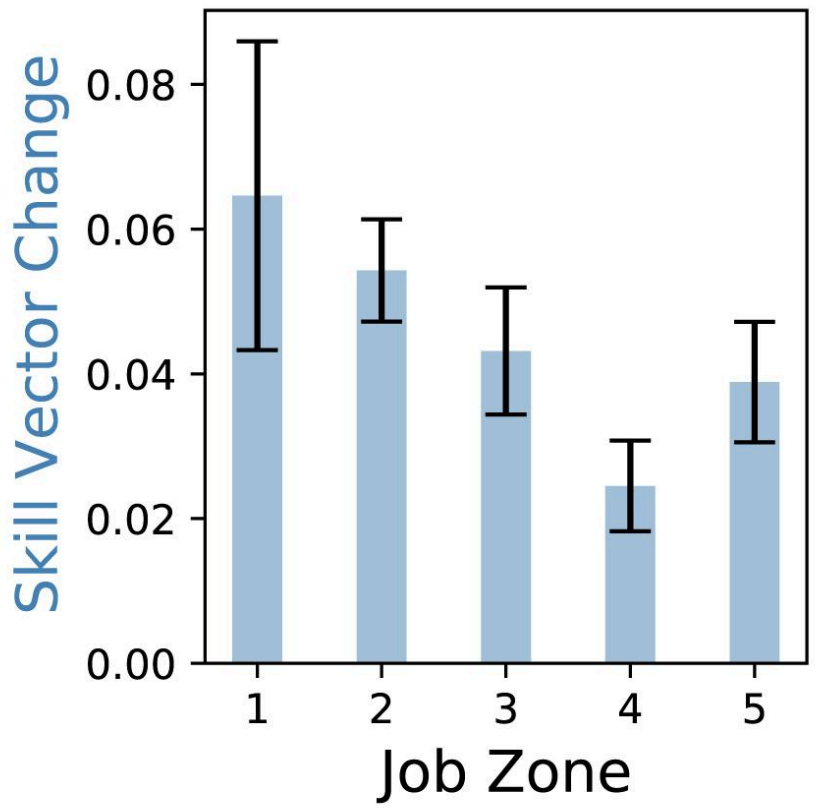

**Supplementary Figure 7. Job zone and skill change:** Occupations have higher average skill content change as job zone increase from 1-4. Each bar denotes the average skill vector change for a group of jobs in a specific job zone labeled on the *x*-axis.

## 4.8 Figure S8. Upskilling of low-skilled jobs towards high-skilled jobs

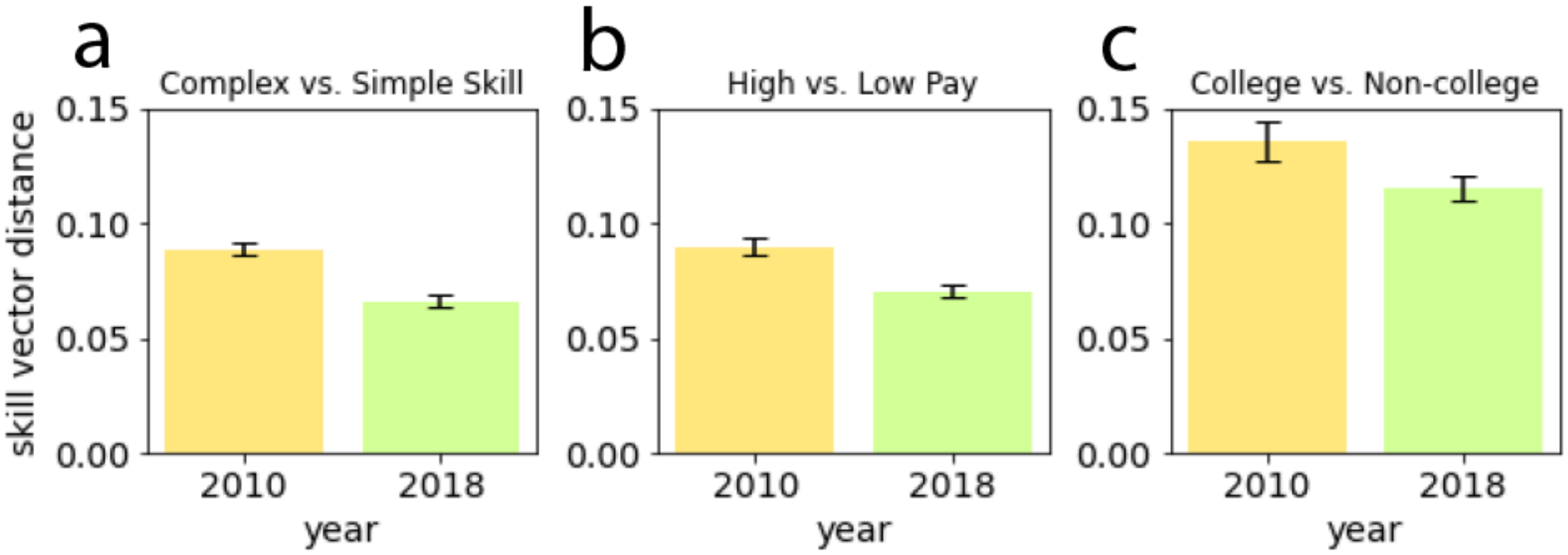

**Supplementary Figure 8. Upskilling of low-skilled jobs towards high-skilled jobs. a.** The skill distance between occupations with relatively complex and simple skills have narrowed from 2010 to 2018. The left yellow bar represents the average skill vector distance between the occupation group with more complex skill requirement (at or above median core skill number across all occupations) and the group of occupations with simpler skill requirements in 2010. The right green bar represents such distance in 2018. **b.** The skill distance between occupations with relatively higher and lower pay have narrowed from 2010 to 2018. The left yellow bar represents the average skill vector distance between the occupation group with higher pay (at or above median in the distribution of annual median salary across all occupations) and the group of occupations with lower pay in 2010. The right green bar represents such distance in 2018. **c.** The skill distance between occupations with college and above vs. non-college degree requirements have narrowed from 2010 to 2018. The left yellow bar represents the average skill vector distance between the occupation group with college and above education requirement and the group of occupations with non-college degree requirements in 2010. The right green bar represents such distance in 2018.

## 4.9 Figure S9. Distribution of number of added skills by occupation

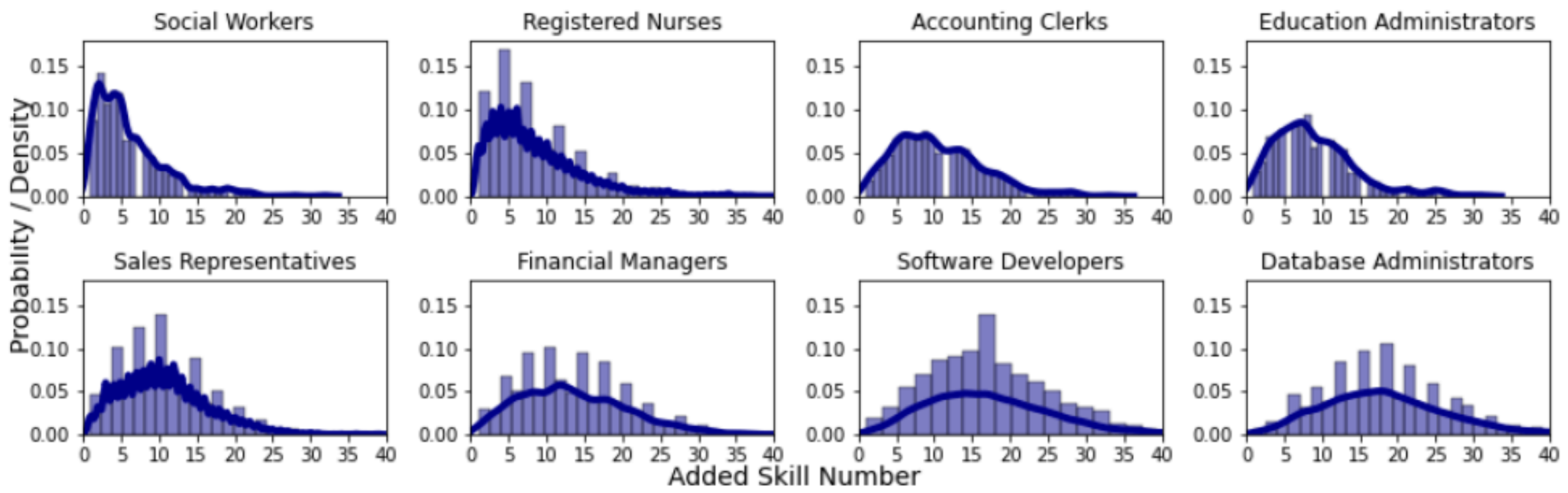

**Supplementary Figure 9. Distribution of number of added skills by occupation.** Each panel is constructed based on all pairs of within-occupation 2010 job posts, $x$ and $y$, where $y$ encompasses all skills required by $x$ plus additional ones, and meanwhile has a higher education degree requirement than $x$. Each panel presents the distribution of the number of added skills in $y$ compared to $x$ from all pairs of $x$ and $y$ under the title occupation.

## 4.10 Figure S10. Education Cost of Marginal Skill Vector Distance between Jobs

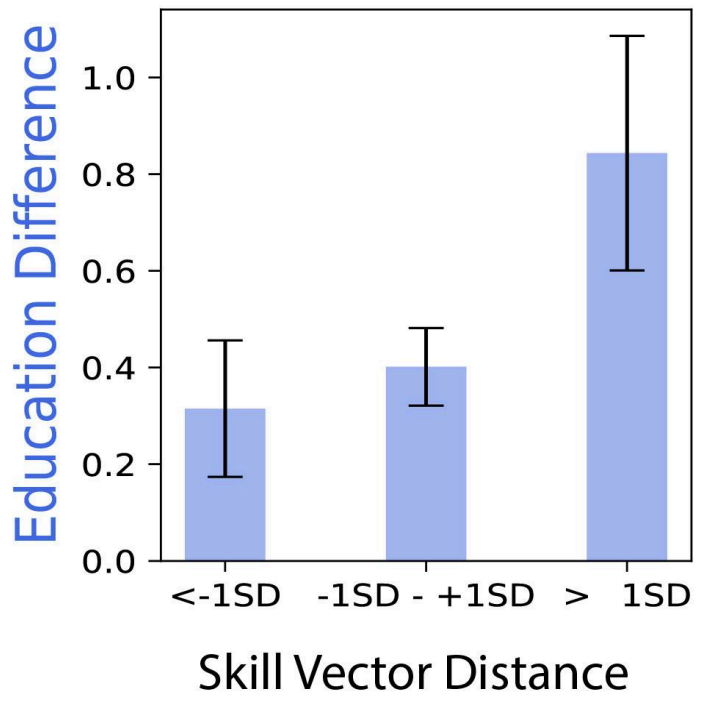

**Supplementary Figure 10. Education Cost of Marginal Skill Vector Distance between Jobs.** Skill vector distance between jobs corresponds to the amount of re-education cost required for workers to move from one distribution of skills to another. The three bars represent the average required education year difference for pairs of job posts with different levels of skill vector distances: the first bar consists of pairs with skill vector distances less than 1 standard deviation below the mean, the second bar on pairs with skill vector distances from 1 standard deviation below the mean to one above the mean, and the third bar includes pairs with skill vector distance larger than 1 standard deviation above the mean.

## 4.11 Figure S11. Unemployment rate and job skill change

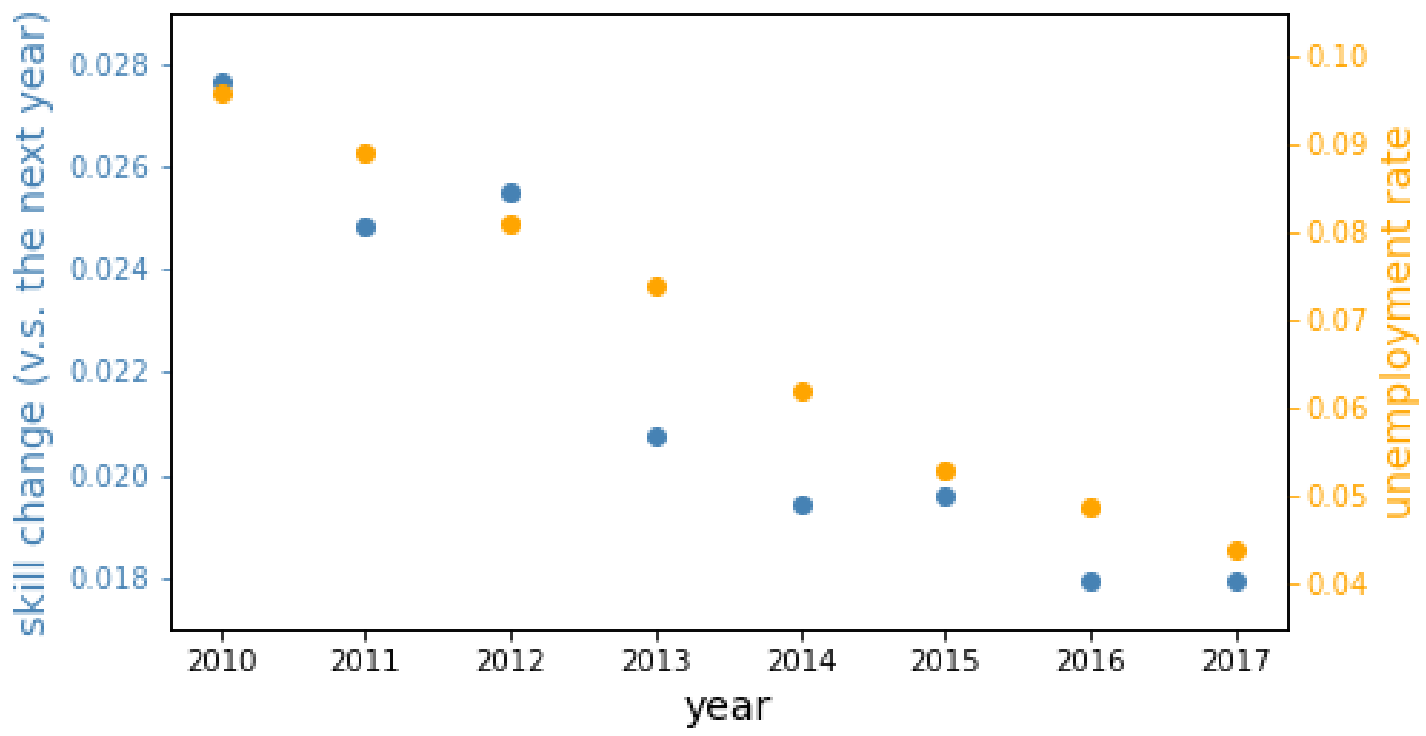

**Supplementary Figure 11. Unemployment rate and job skill change.** The blue dots represent the year-on-year (e.g., 2010-2011, 2011-2012, etc.) average occupational skill change, corresponding to the blue y-axis on the left. The orange dots represent the yearly unemployment rate in the U.S., corresponding to the orange y-axis on the right.

## 5. Supplementary Tables

### 5.1 Table S1. Explaining job skill change variation with skill complexity, measuring change with the most significant skill transition

| | *Dependent variable: Most Significant Skill Transition* | | | | | | | | |
|---|---|---|---|---|---|---|---|---|---|
| | Max | Max | Max | Rank 5% | Rank 5% | Rank 5% | Rank 10% | Rank 10% | Rank 10% |
| Skill Complexity | -0.010*** | | | -0.053*** | | | -0.062*** | | |
| | (0.002) | | | (0.002) | | | (0.002) | | |
| Log Annual Pay | | -0.021*** | | | -0.053*** | | | -0.069*** | |
| | | (0.006) | | | (0.007) | | | (0.007) | |
| Education | | | -0.004*** | | | -0.009*** | | | -0.010*** |
| | | | (0.001) | | | (0.001) | | | (0.001) |
| Observations | 721 | 721 | 721 | 721 | 721 | 721 | 721 | 721 | 721 |
| $R^2$ | 0.025 | 0.019 | 0.025 | 0.486 | 0.081 | 0.072 | 0.536 | 0.113 | 0.086 |
| Adjusted $R^2$ | 0.024 | 0.018 | 0.024 | 0.485 | 0.080 | 0.071 | 0.535 | 0.112 | 0.085 |
| Residual Std. Error | 0.057 (df=719) | 0.057 (df=719) | 0.057 (df=719) | 0.050 (df=719) | 0.067 (df=719) | 0.068 (df=719) | 0.053 (df=719) | 0.073 (df=719) | 0.074 (df=719) |
| F Statistic | 18.775*** (df=1; 719) | 14.079*** (df=1; 719) | 18.379*** (df=1; 719) | 679.363*** (df=1; 719) | 63.626*** (df=1; 719) | 55.877*** (df=1; 719) | 829.453*** (df=1; 719) | 91.385*** (df=1; 719) | 67.906*** (df=1; 719) |
| Note: | | | | | | | | | *p<0.05; **p<0.01; ***p<0.001 |

### 5.2 Table S2. Example of job posts pairs used for predicting re-education cost

| | Job Ad *i* | Job Ad j | Job Ad *i* | Job Ad *j* |
|---|---|---|---|---|
| BGTJobId | 311043099 | 311477284 | 311765112 | 311777227 |
| Title | Administrative Assistant | Administrative Assistant III | Registered Nurse Signature Suites Vip | Registered Nurse Interventional Radiology N |
| Education Years | 14 | 16 | 14 | 16 |
| Common Skills | Communication Skills<br>Microsoft Excel<br>Microsoft Powerpoint<br>Administrative Support<br>Microsoft Word | Communication Skills<br>Microsoft Excel<br>Microsoft Powerpoint<br>Administrative Support<br>Microsoft Word | Budgeting<br>Patient Care<br>Process Improvement<br>Research<br>Teamwork/Collaboration<br>Customer Contact<br>Life Support<br>Planning | Budgeting<br>Patient Care<br>Process Improvement<br>Research<br>Teamwork/Collaboration<br>Customer Contact<br>Life Support<br>Planning |
| Additional skill | | Customer Information Control System (CICS)<br>Direct Mail<br>Newsletters<br>Portal Tools<br>Siebel CRM | | Advanced Cardiac Life Support (ACLS)<br>Catheterization<br>Catheterization Laboratory (CATH LAB)<br>Critical Care<br>Interventional Radiology (IR) |
| Skill Distance | 0.17 | | 0.15 | |

## 5.3 Table S3. Predicting education year difference with skill distance

| | *Dependent variable: Education Year Difference* | | |
|---|---|---|---|
| | Job Post Unit | Occupation Aggregated Unit | Occupation Aggregated Unit |
| Skill Distance | 0.741*** | 1.839*** | |
| | (0.007) | (0.352) | |
| Skill Distance Z-score | | | 0.172*** |
| | | | (0.033) |
| Observations | 6,585,524 | 1,257 | 1,257 |
| $R^2$ | 0.002 | 0.021 | 0.021 |
| Adjusted $R^2$ | 0.002 | 0.020 | 0.020 |
| Residual Std. Error | 2.038 (df=6585522) | 1.167 (df=1255) | 1.167 (df=1255) |
| F Statistic | 11524.516*** (df=1; 6585522) | 27.230*** (df=1; 1255) | 27.230*** (df=1; 1255) |
| Note: | | | *p<0.05; **p<0.01; ***p<0.001 |

### 5.4 Table S4. Explaining occupation skill change variation with skill complexity

| | Dependent variable: Occupation skill change | | | | | |
|---|---|---|---|---|---|---|
| | Model 1 | Model 2 | Model 3 | Model 4 | Model 5 | Model 6 |
| Skill Complexity | -0.012*** | | | -0.012*** | | |
| | (0.001) | | | (0.001) | | |
| Log Annual Pay | | -0.007*** | | | -0.005*** | |
| | | (0.001) | | | (0.001) | |
| Education | | | -0.001*** | | | -0.002*** |
| | | | (0.000) | | | (0.000) |
| Log Emp. Concentr. | | | | 0.001 | 0.013*** | 0.015*** |
| | | | | (0.001) | (0.001) | (0.001) |
| Job Similarity Increase | | | | -0.030* | -0.027* | -0.022 |
| | | | | (0.012) | (0.013) | (0.013) |
| Job Similarity | | | | -0.027*** | -0.022** | -0.014 |
| | | | | (0.007) | (0.007) | (0.007) |
| Observations | 668 | 668 | 668 | 668 | 668 | 668 |
| $R^2$ | 0.300 | 0.037 | 0.033 | 0.319 | 0.178 | 0.199 |
| Adjusted $R^2$ | 0.299 | 0.035 | 0.032 | 0.315 | 0.173 | 0.194 |
| Residual Std. Error | 0.046 (df=666) | 0.058 (df=666) | 0.058 (df=666) | 0.045 (df=663) | 0.052 (df=663) | 0.052 (df=663) |
| F Statistic | 284.882*** (df=1; 666) | 25.304*** (df=1; 666) | 22.842*** (df=1; 666) | 77.807*** (df=4; 663) | 35.959*** (df=4; 663) | 41.258*** (df=4; 663) |

Note: *p<0.05; **p<0.01; ***p<0.001

Log Emp. Concentr. refers to the natural log of HHI (employer concentration).Job similarity measures how similar job posts are within the same firm, market, occupation in a given year.

## 5.5 Table S5. Explaining skill change variation with employer and market size, within firm-region-occupation change

| | *Dependent variable: Firm-region-occupation skill change* | | | | |
|---|---|---|---|---|---|
| | Model 1 | Model 2 | Model 3 | Model 4 | Model 5 |
| Employer Size | -0.004*** | | -0.003*** | | -0.002† |
| | (0.001) | | (0.001) | | (0.001) |
| CZ Market Size | | -0.008*** | | -0.022*** | -0.022*** |
| | | (0.001) | | (0.002) | (0.002) |
| Log Emp. Concentr. | | | -0.005*** | -0.021*** | -0.021*** |
| | | | (0.001) | (0.002) | (0.002) |
| Job Similarity Increase | | | 0.099*** | 0.102*** | 0.099*** |
| | | | (0.015) | (0.015) | (0.015) |
| Occ. FE | Yes | Yes | Yes | Yes | Yes |
| Observations | 4497 | 4497 | 4497 | 4497 | 4497 |
| $R^2$ | 0.113 | 0.116 | 0.124 | 0.150 | 0.151 |
| Adjusted $R^2$ | 0.088 | 0.091 | 0.099 | 0.126 | 0.126 |
| Residual Std. Error | 0.084 (df=4374) | 0.084 (df=4374) | 0.084 (df=4372) | 0.083 (df=4372) | 0.083 (df=4371) |
| F Statistic | 4.569*** (df=122; 4374) | 4.700*** (df=122; 4374) | 4.987*** (df=124; 4372) | 6.227*** (df=124; 4372) | 6.208*** (df=125; 4371) |

Note: †p<0.07; *p<0.05; **p<0.01; ***p<0.001

CZ refers to commuting zone. Log Emp. Concentr. refers to the natural log of HHI (employer concentration). Job similarity Increase measures the difference between 2018 and 2010 within-firm-cz-occupation job homogeneity level.

### 5.6 Table S6. Explaining skill change variation with employer and market size, firm-region overall change

| | *Dependent variable: Firm-region skill change* | | | | |
|---|---|---|---|---|---|
| | Model 1 | Model 2 | Model 3 | Model 4 | Model 5 |
| Employer Size | -0.001** | | -0.001* | | -0.001* |
| | (0.000) | | (0.000) | | (0.000) |
| CZ Market Size | | -0.003*** | | -0.019*** | -0.019*** |
| | | (0.001) | | (0.004) | (0.004) |
| Log Emp. Concentr. | | | 0.012*** | -0.066*** | -0.066*** |
| | | | (0.003) | (0.015) | (0.015) |
| Job Similarity Increase | | | -0.005 | -0.004 | -0.001 |
| | | | (0.018) | (0.018) | (0.018) |
| Observations | 13141 | 13141 | 13141 | 13141 | 13141 |
| $R^2$ | 0.001 | 0.002 | 0.002 | 0.004 | 0.004 |
| Adjusted $R^2$ | 0.000 | 0.002 | 0.002 | 0.004 | 0.004 |
| Residual Std. Error | 0.109 (df=13139) | 0.109 (df=13139) | 0.109 (df=13137) | 0.109 (df=13137) | 0.108 (df=13136) |
| F Statistic | 6.731** (df=1; 13139) | 30.511*** (df=1; 13139) | 8.941*** (df=3; 13137) | 16.793*** (df=3; 13137) | 13.776*** (df=4; 13136) |
| Note: | | | | | *p<0.05; **p<0.01; ***p<0.001 |

CZ refers to commuting zone. Log Emp. Concentr. refers to the natural log of HHI (employer concentration). Job similarity Increase measures the difference between 2018 and 2010 within-firm-cz-occupation job homogeneity level.

## 5.7 Table S7. Explaining occupation skill change variation with skill complexity, different job content scope and skill weights

| | *Dependent variable: Occupation skill change* | | |
|---|---|---|---|
| | All Skills | Top 50% Skills | Top 25% Skills |
| Skill Complexity | -0.006*** | -0.007*** | -0.007*** |
| | (0.001) | (0.001) | (0.001) |
| Log Emp. Concentr. | 0.003* | 0.003* | 0.003* |
| | (0.002) | (0.002) | (0.002) |
| Job Similarity Increase | -0.050*** | -0.051*** | -0.053*** |
| | (0.012) | (0.012) | (0.013) |
| Job Similarity | -0.029*** | -0.030*** | -0.031*** |
| | (0.007) | (0.007) | (0.007) |
| Observations | 668 | 668 | 668 |
| $R^2$ | 0.171 | 0.175 | 0.183 |
| Adjusted $R^2$ | 0.166 | 0.170 | 0.178 |
| Residual Std. Error | 0.031 (df=663) | 0.032 (df=663) | 0.036 (df=663) |
| F Statistic | 34.129*** (df=4; 663) | 35.044*** (df=4; 663) | 37.004*** (df=4; 663) |
| Note: | | | *p<0.05; **p<0.01; ***p<0.001 |

Log Emp. Concentr. refers to the natural log of HHI (employer concentration).Job similarity measures how similar job posts are within the same firm, market, occupation in a given year.

## 5.8 Table S8. Explaining skill change variation with employer and market size (within firm-region-occupation change), different job content scope and skill weights

| | *Dependent variable: Firm-region-occupation skill change* | | |
|---|---|---|---|
| | All Skills | Top 50% Skills | Top 25% Skills |
| Employer Size | -0.005*** | -0.006*** | -0.006*** |
| | (0.000) | (0.000) | (0.000) |
| CZ Market Size | -0.009*** | -0.011*** | -0.013*** |
| | (0.000) | (0.000) | (0.000) |
| Log Emp. Concentr. | -0.007*** | -0.010*** | -0.013*** |
| | (0.000) | (0.000) | (0.000) |
| Job Similarity Increase | 0.000 | 0.003* | 0.009*** |
| | (0.001) | (0.001) | (0.001) |
| Occ. FE | Yes | Yes | Yes |
| Observations | 283151 | 283151 | 283151 |
| $R^2$ | 0.138 | 0.146 | 0.155 |
| Adjusted $R^2$ | 0.136 | 0.144 | 0.153 |
| Residual Std. Error | 0.104 (df=282484) | 0.108 (df=282484) | 0.116 (df=282484) |
| F Statistic | 68.046*** (df=666; 282484) | 72.578*** (df=666; 282484) | 77.927*** (df=666; 282484) |

Note: *p<0.05; **p<0.01; ***p<0.001

CZ refers to commuting zone. Log Emp. Concentr. refers to the natural log of HHI (employer concentration). Job similarity Increase measures the difference between 2018 and 2010 within-firm-cz-occupation job homogeneity level.

### 5.9 Table S9. Explaining skill change variation with employer and market size ( firm-region overall change), different job content scope and skill weights

| | *Dependent variable: Firm-region skill change* | | |
|---|---|---|---|
| | All Skills | Top 50% Skills | Top 25% Skills |
| Employer Size | -0.009*** | -0.009*** | -0.010*** |
| | (0.000) | (0.000) | (0.000) |
| CZ Market Size | -0.023*** | -0.027*** | -0.030*** |
| | (0.001) | (0.001) | (0.001) |
| Log Emp. Concentr. | -0.058*** | -0.066*** | -0.068*** |
| | (0.004) | (0.004) | (0.005) |
| Job Similarity Increase | -0.146*** | -0.141*** | -0.132*** |
| | (0.004) | (0.004) | (0.005) |
| Observations | 143934 | 143934 | 143934 |
| $R^2$ | 0.073 | 0.073 | 0.077 |
| Adjusted $R^2$ | 0.073 | 0.073 | 0.077 |
| Residual Std. Error | 0.134 (df=143929) | 0.137 (df=143929) | 0.146 (df=143929) |
| F Statistic | 2838.982*** (df=4; 143929) | 2838.661*** (df=4; 143929) | 2993.067*** (df=4; 143929) |
| Note: | | | *p<0.05; **p<0.01; ***p<0.001 |

CZ refers to commuting zone. Log Emp. Concentr. refers to the natural log of HHI (employer concentration). Job similarity Increase measures the difference between 2018 and 2010 within-firm-cz-occupation job homogeneity level.